\documentclass[a4paper,draft]{agujournal2019}
\usepackage[T1]{fontenc}
\usepackage[latin9]{inputenc}
\usepackage{units}
\usepackage{url}
\usepackage{amsbsy}
\usepackage{amstext}
\usepackage{graphicx}
\usepackage{rotfloat}
\usepackage{esint}
\usepackage[authoryear]{natbib}
\usepackage[unicode=true,pdfusetitle,
 bookmarks=true,bookmarksnumbered=false,bookmarksopen=false,
 breaklinks=true,pdfborder={0 0 0},pdfborderstyle={},backref=false,colorlinks=false]
 {hyperref}
\hypersetup{
 final}

\makeatletter

\pdfpageheight\paperheight
\pdfpagewidth\paperwidth

\journalname{Journal of Geophysical Research: Space Physics}

\usepackage{caption}
\makeatother

\begin{document}

\title{Auroral, Ionospheric and Ground Magnetic Signatures of Magnetopause
Surface Modes}

\authors{M. O. Archer, \affil{1}\\
M. D. Hartinger, \affil{2}\\
L. Rast\"{a}tter, \affil{3}\\
D. J. Southwood, \affil{1}\\
M. Heyns, \affil{1}\\
J. W. B. Eggington, \affil{1}\\
A. N. Wright, \affil{4}\\
F. Plaschke, \affil{5}\\
and X. Shi \affil{6,7}}

\affiliation{1}{Space and Atmospheric Physics Group, Department of Physics, Imperial
College London, London, UK.}

\affiliation{2}{Space Science Institute, Boulder, Colorado, USA.}

\affiliation{3}{NASA Goddard Space Flight Center, Greenbelt, Maryland, USA.}

\affiliation{4}{Department of Mathematics and Statistics, University of St Andrews,
St Andrews, UK.}

\affiliation{5}{ Institut für Geophysik und extraterrestrische Physik, TU Braunschweig,
Braunschweig, Germany.}

\affiliation{6}{Department of Electrical and Computer Engineering, Virginia Polytechnic
Institute and State University, Blacksburg, Virginia, USA.}

\affiliation{7}{High Altitude Observatory, National Center for Atmospheric Research,
Boulder, Colorado, USA.}

\correspondingauthor{Martin Archer}{m.archer10@imperial.ac.uk}

\begin{keypoints}

\item Theory and global simulations of magnetopause surface waves'
effects on the aurorae, ionosphere, and ground magnetic field are
investigated

\item We predict poleward-moving periodic aurora, convection vortices,
and ground pulsations, with larger latitudinal scales than Alfv\'{e}n
modes

\item Amplitudes of all signals peak near the projection of the inner/equatorward
edge of the magnetopause rather than the open--closed boundary

\end{keypoints}
\begin{abstract}
Surface waves on Earth\textquoteright s magnetopause have a controlling
effect upon global magnetospheric dynamics. Since spacecraft provide
sparse \textit{in situ} observation points, remote sensing these modes
using ground-based instruments in the polar regions is desirable.
However, many open conceptual questions on the expected signatures
remain. Therefore, we provide predictions of key qualitative features
expected in auroral, ionospheric, and ground magnetic observations
through both magnetohydrodynamic theory and a global coupled magnetosphere-ionosphere
simulation of a magnetopause surface eigenmode. These show monochromatic
oscillatory field-aligned currents, due to both the surface mode and
its non-resonant Alfv\'{e}n coupling, are present throughout the
magnetosphere. The currents peak in amplitude at the equatorward edge
of the magnetopause boundary layer, not the open-closed boundary as
previously thought. They also exhibit slow poleward phase motion rather
than being purely evanescent. We suggest the upward field-aligned
current perturbations may result in periodic auroral brightenings.
In the ionosphere, convection vortices circulate the poleward moving
field-aligned current structures. Finally, surface mode signals are
predicted in the ground magnetic field, with ionospheric Hall currents
rotating perturbations by approximately (but not exactly) $90^{\circ}$
compared to the magnetosphere. Thus typical dayside magnetopause surface
modes should be strongest in the East-West ground magnetic field component.
Overall, all ground-based signatures of the magnetopause surface mode
are predicted to have the same frequency across $L$-shells, amplitudes
that maximise near the magnetopause's equatorward edge, and larger
latitudinal scales than for field line resonance. Implications in
terms of ionospheric Joule heating and geomagnetically induced currents
are discussed.
\end{abstract}

\section*{Plain Language Summary}

Waves on the boundary of the magnetosphere, the magnetic shield established
by the interplay of the solar wind with Earth's magnetic field, play
a controlling role on energy flow into our space environment. While
these waves can be observed as they pass over satellites in orbit,
due to the small number of suitable satellites available it would
be helpful to be able to detect these waves from the surface of the
Earth with instruments that measure the northern/southern lights,
motion of the top of our atmosphere, or magnetic field on the ground.
However, we do not currently understand what the signs of these waves
should look like in such instruments. In this paper we develop theory
and use computer simulations of these boundary waves to predict key
features one might expect to measure from the ground. Based on these
predictions, we also discuss how the waves might contribute to the
hazards of space weather.

\section{Introduction}

The interaction between the solar wind and Earth's magnetosphere results
in a zoo of dynamical plasma waves. Those with wavelengths comparable
to the size of the magnetosphere are well described by magnetohydrodynamics
(MHD) and due to their corresponding frequencies, $\sim0.1\text{--}100\,\mathrm{mHz}$
\citep{Jacobs1964}, are known as ultra-low frequency (ULF) waves.
ULF waves play important roles in space weather processes such as
substorms \citep[e.g.][]{Kepko1999}, wave-wave \citep[e.g.][]{Li2011}
and wave-particle \citep[e.g.][]{Turner2012} interactions, magnetosphere-ionosphere
(MI) coupling \citep[e.g.][]{Keiling2009}, and geomagnetically induced
currents \citep[e.g.][]{Heyns2021}. In addition to familiar Alfv\'{e}n
and fast/slow magnetosonic body MHD waves (those which may freely
propagate through plasma volumes), sharp discontinuities separating
regions with different physical parameters, such as the magnetopause
and plasmapause, allow for collective modes -- surface waves \citep{Kruskal1954,Goedbloed1971,Chen1974}.
Surface modes lead to mass, momentum, and energy transport across
the boundary, consequently manifesting a controlling effect on global
magnetospheric wave dynamics \citep[e.g.][]{Kivelson1995}. As with
the body waves, theory behind surface waves has largely been developed
in simplified box model magnetospheres, typically with homogeneous
half-spaces. Within these the surface mode is inherently compressional,
being described by two evanescent fast magnetosonic waves (one in
each half-space such that perturbations decay with distance from the
boundary) that are joined by boundary conditions which ensure pressure
balance and continuity of normal displacement across the interface
\citep{Pu1983,Plaschke2011}. On each side the magnetosonic relation

\begin{equation}
k_{n}^{2}=-k_{\perp}^{2}-k_{\parallel}^{2}+\frac{\omega^{4}}{\omega^{2}v_{A}^{2}+c_{s}^{2}\left(\omega^{2}-k_{\parallel}^{2}v_{A}^{2}\right)}\label{eq:magnetosonic-dispersion}
\end{equation}
holds, where $n$ represents the direction normal to the discontinuity,
and $v_{A}$ and $c_{s}$ are the Alfv\'{e}n and sound speeds respectively
(see Notation). Under incompressibility, the last term of equation~\ref{eq:magnetosonic-dispersion}
may be neglected and the normal wavenumber is imaginary. In contrast,
if this assumption is not valid or if waves are damped/unstable then
$k_{n}$ may be complex, exhibiting both evanescence and normal phase
motion \citep[hereafter A21]{Pu1983,Archer2021}. The dispersion relation
for incompressible surface waves in a box model can be analytically
solved. Applied to the magnetopause, for zero magnetic shear across
boundary (equivalent to northward interplanetary magnetic field; IMF)
and no background flows, it is \citep{Plaschke2011}
\begin{equation}
\omega=k_{\parallel}\sqrt{\frac{B_{msp}^{2}+B_{msh}^{2}}{\mu_{0}\left(\rho_{msp}+\rho_{msh}\right)}}\approx k_{\parallel}\frac{B_{msp}}{\sqrt{\mu_{0}\rho_{msh}}}
\end{equation}
where $msp$ refers to the magnetosphere and $msh$ the magnetosheath.
The boundary conditions of closed magnetic field lines at the northern
and southern ionospheres impose quantised wavelengths along the field
\citep{Chen1974}, forming an eigenmode of the system. On the dayside,
where magnetosheath flows are smaller, this magnetopause surface eigenmode
(MSE) is expected to occupy frequencies below $2\,\mathrm{mHz}$ \citep{Plaschke2009,Archer2015}.
Such low eigenfrequencies are a result of the combination of magnetic
fields and densities from both sides of the magnetopause, making it
the lowest frequency magnetospheric normal mode and highly penetrating.
However, in the flanks the faster magnetosheath velocities are expected
to dictate the wave frequency \citep{Plaschke2011,Kozyreva2019},
rather than the extent of the field lines, yielding shorter wavelengths
and periods. In the absence of continuous driving or instabilities,
surface modes on a boundary of finite thickness are strongly damped,
with this thought to be primarily due to mode conversion to Alfv\'{e}n
waves and spatial phase mixing within the boundary layer rather than
dissipation in the ionosphere or due to the presence of the ionosphere-Earth
boundary \citep{Chen1974}.

Magnetopause surface modes may be excited by several driving processes,
either external or internal to the magnetosphere. External mechanisms
include upstream (solar wind, foreshock, or magnetosheath) pressure
variations, which may be either quasi-periodic \citep[e.g.][]{Sibeck1989}
or impulsive \citep[e.g.][]{Shue2009}, and the Kelvin-Helmholtz instability
(KHI) due to velocity shears \citep[e.g.][]{Fairfield2000}. Internal
processes, such as the drift mirror instability, can generate compressional
ULF waves within the low- and high-latitude magnetospheric boundary
layer \citep{Constantinescu2009,Nykyri2021}, which may also lead
to surface wave growth at the magnetopause. There has been much evidence
of magnetopause surface waves from spacecraft observations, particularly
in the magnetospheric flanks where KHI-generated waves are thought
to be prevalent \citep[e.g.][]{Southwood1968,Kavosi2015}. However,
only recently was MSE, as proposed by \citet{Chen1974}, discovered
through multi-spacecraft observations on the dayside magnetosphere
following impulsive external driving \citep{Archer2019}.

Understanding the fundamental properties and potential impacts of
magnetopause surface modes within a realistic magnetospheric environment
has necessitated the use of global MHD simulations \citep[e.g.][henceforth H15]{Claudepierre2008,Hartinger2015}.
These have revealed surface waves might lead the entire magnetosphere
to oscillate at the surface mode frequency by coupling to body MHD
waves such as Alfv\'{e}nic field line resonance (FLR) or fast magnetosonic
waveguide modes (WM; \citealp{Merkin2013}; A21). Confirming such
a global system response is challenging with \textit{in situ} spacecraft
observations. For any particular event, only a few observation points
are available from current missions (e.g. Cluster, THEMIS, MMS). Statistical
studies are also difficult due to the highly variable conditions present
throughout geospace, which influence the properties of ULF waves \citep{Archer2015,Archer2015a}.
On the other hand, ground-based instruments such as all-sky imagers
\citep[e.g.][]{Donovan2006,Rae2012}, radar \citep[e.g.][]{Walker1979,Nishitani2019},
and ground magnetometers \citep[e.g.][]{Mathie1999,Gjerloev2009}
provide good coverage of the near-Earth signatures of ULF waves. While
they offer the possibility of remote sensing the magnetopause surface
mode, we need to understand how its energy couples through the intervening
regions. The theory behind ionospheric and ground effects of ULF waves
has focused on the Alfv\'{e}n mode \citep[e.g.][]{hughes74nature,Hughes1976},
whereas the surface mode is fundamentally compressional. Several open
conceptual questions about the nature of surface modes in these regions
remain. It is also difficult to confidently distinguish with ground-based
instruments between surface waves, either on the magnetopause \citep{Kivelson1991,Glassmeier1992,Glassmeier1992b}
or low latitude boundary layer \citep{Sibeck1990,lyatsky97}, and
propagating body waves near these boundaries \citep{Tamao1964a,Tamao1964b,Araki1988,slinker99},
especially as surface waves may excite secondary body waves (\citealp{Southwood1974};
A21).

It is not clear whether magnetopause surface waves are expected to
directly affect the ionosphere. \citet{Kivelson1988} consider the
currents and boundary conditions associated with MHD waves in a box
model. They argue surface waves affect the ionosphere only across
the thin transition layer. This work was, however, applied to the
plasmapause, avoiding the complication with the magnetopause that
adjacent magnetosheath field lines do not terminate in the polar cap
\citep{Kozyreva2019}. Other theoretical models have focused on closed
field lines Earthward of the boundary, considering field-aligned current
(FAC) generation due to coupling between the compressional and Alfv\'{e}n
modes in an inhomogeneous/curvilinear magnetosphere \citep{Sibeck1990,southwood90,southwood91}.
The models all predict FACs communicate (tailward travelling) magnetopause
disturbances to the ionosphere, resulting in so-called travelling
convection vortices (TCVs). These were first inferred from ground
magnetometer observations \citep{FriisChristensen1988} and can also
be observed directly through radar observations \citep[e.g.][]{Bristow1995}.
Discrete auroral emission might also result from precipitating electrons
which carry these FACs \citep{Greenwald1980}. The models, however,
do not make predictions about the magnetopause surface mode directly.
In particular, they circumvent the question of how the ionosphere
is affected by field lines within (and close to) the boundary layer.
Furthermore, auroral brightenings and TCVs are expected for any magnetospheric
process which results in FACs, e.g. field line resonance \citep{Greenwald1980},
hence predictions of how to distinguish effects caused by surface
waves and other mechanisms are required.

The direct ground magnetic field signatures of surface waves are also
poorly understood, even during confirmed case studies from \textit{in
situ} spacecraft observations \citep{Archer2019,He2020}. \citet{Kivelson1988}
suggest the surface mode may be screened from the ground due to the
thin ionospheric region affected, similar to with small wavelength
Alfv\'{e}n modes \citep{Hughes1976}. They conclude the magnetic
signal on the ground might be similar to that in the magnetosphere,
i.e. not rotated by $\sim90^{\circ}$ as Alfv\'{e}n waves are \citep{hughes74,hughes74nature}.
However, vortical ground magnetic signals are often observed by high-latitude
magnetometer networks, being associated with TCVs \citep[e.g.][]{Glassmeier1992,Glassmeier1992b,Hwang2022}.
This potentially calls the theoretical prediction into question or
signals intermediate Alfv\'{e}n waves may be involved.

Finally, it is not clear where auroral, ionospheric, and ground magnetic
signals of magnetopause surface waves should map to. Intuitively one
might expect them around the open-closed boundary (OCB) of magnetic
field lines. \citet{Kozyreva2019} suggest that short-lived quasi-periodic
motions of the OCB in auroral keograms and ground magnetic oscillations
near the OCB with large latitudinal scales and similar periodicities
across $L$-shells might distinguish magnetopause surface modes from
the Alfv\'{e}n continuum, presenting potential case studies. However,
Pc5-6 band (periods $\sim3\text{--}15\,\mathrm{min}$) oscillations
in ground magnetometer data have been shown to peak systematically
equatorward of optical and ionospheric proxies for the cusp OCB by
$1\text{--}3^{\circ}$ \citep{Pilipenko2017,Pilipenko2018,Kozyreva2019}.
In the absence of conjugate space-based observations, conclusions
have been mixed over whether these results relate to MSE and what
the implications are for its excitation efficiency.

To resolve these open questions, we employ MHD theory and a global
MI-coupling simulation of MSE. Since MSE are the lowest frequency
normal mode of the magnetosphere, they allow us to better understand
the direct effects of surface waves on the dayside aurorae, ionosphere,
and ground magnetic field without the complications of secondary coupled
wave modes. We aim to detail the physical processes that lead to these
signatures, yielding specific qualitative predictions that might enable
crucial remote-sensing observations of magnetopause surface modes
in the future.

\section{Box Model Theory\label{sec:Box-Model-Theory}}

We first consider a box model magnetosphere. These straighten the
geomagnetic field lines into a uniform field bounded by northern and
southern ionospheres \citep{Radoski1971,Southwood1974}. While this
is an unrealistic simplification \citep[e.g.][]{Kozyreva2019}, it
is used here to gain initial qualitative insight.

\subsection{Method}

We use the same model setup as \citet{Plaschke2011}, who derived
the magnetospheric signatures of incompressible MSE. The model equilibrium
consists of two uniform half-spaces, the magnetosheath and magnetosphere,
separated by the magnetopause discontinuity at $x=0$. The geomagnetic
field is confined to $x>0$ and points in the $z$-direction. Thus
close to the MI-interface, $x$ is directed equatorward and $y$ is
westward. \citet{Plaschke2011} derived the currents associated with
surface waves in a box model by finding solutions on closed field
lines and requiring continuity of pressure and normal velocity across
the boundary. They showed the currents are sinusoidal and contained
within the infinitesimally thin boundary, i.e. the last shell of closed
field lines, which we refer to as magnetopause currents. These magnetopause
currents have field-aligned components, particularly at the MI-interface.
Fast magnetosonic waves are not expected to have FACs in infinite
uniform media, only Alfv\'{e}n waves lead to these, hence generally
fast magnetosonic modes are not expected to couple to the ionosphere.
The surface mode, however, is unique as a fast mode which supports
FACs at the interface of two uniform half-spaces due to the nonuniformity
at this location. A complementary view considers amplitudes of the
different MHD modes through the divergence and curl of electric field
perturbations $\delta\mathbf{E}$ \citep{Yoshikawa1996,yoshikawa00},
where $\left(\nabla\times\delta\mathbf{E}\right)_{\parallel}$ gives
the fast/compressional mode and $\nabla\cdot\delta\mathbf{E}$ yields
the Alfv\'{e}n/shear mode. Applying this to the \citet{Plaschke2011}
analytic solutions reveals that in the two uniform half-spaces the
surface wave is purely compressional ($\nabla\cdot\delta\mathbf{E}=0$),
whereas inside the boundary layer there are non-zero amplitudes for
both shear and compressional modes (via Gauss' and Stokes' theorems
respectively).

\citet{Plaschke2011} suggested the surface waves' FACs at the MI-interface
might close in the ionosphere. We, therefore, extend their model to
incorporate a finite conductivity thin-shell ionosphere using the
electrostatic MI-coupling method \citep{Wolf1975,Goodman1995,Janhunen1998,Ridley2004},
valid since surface waves occupy such low frequencies \citep{lotko04}.
This works by determining the disturbance ionospheric potential $\delta\psi_{isp}$
through current continuity, which for the northern hemisphere is given
by

\begin{equation}
\begin{array}{rl}
\delta j_{r} & =\nabla_{\perp}\cdot\left(\boldsymbol{\Sigma}\cdot\nabla\delta\psi_{isp}\right)_{\perp}\\
 & =\nabla_{\perp}\cdot\left[\left(\begin{array}{cc}
\Sigma_{P} & -\Sigma_{H}\\
\Sigma_{H} & \Sigma_{P}
\end{array}\right)\cdot\nabla\delta\psi_{isp}\right]\\
 & =\Sigma_{P}\left(\frac{\partial^{2}}{\partial x^{2}}+\frac{\partial^{2}}{\partial y^{2}}\right)\delta\psi_{isp}
\end{array}\label{eq:j-continuity}
\end{equation}
where $\delta j_{r}$ are the vertical currents (pointing radially/upwards)
at the MI-interface from \citet{Plaschke2011}, and $\boldsymbol{\Sigma}$
denotes the height-integrated conductivity tensor consisting of Pedersen
(P) and Hall (H) conductances, both assumed to be uniform. Equation~\ref{eq:j-continuity}
is solved numerically using the 2D Laplacian's Green's function
\begin{equation}
\delta\psi_{isp}\left(x,y\right)=\iint dx^{\prime}dy^{\prime}\frac{\delta j_{r}\left(x^{\prime},y^{\prime}\right)}{\Sigma_{P}}\frac{\ln\left(\sqrt{(x-x^{\prime})^{2}+(y-y^{\prime})^{2}}\right)}{2\pi}
\end{equation}
from which the ionospheric electric field and currents are determined.
Finally, magnetic field perturbations at some location $\mathbf{r}$
are calculated using the Biot-Savart law

\begin{equation}
\delta\mathbf{B}\left(\mathbf{r}\right)=\frac{\mu_{0}}{4\pi}\iiint d^{3}\mathbf{r}^{\prime}\frac{\delta\mathbf{j}\left(\mathbf{r}^{\prime}\right)\times\left(\mathbf{r}-\mathbf{r}^{\prime}\right)}{\left|\left(\mathbf{r}-\mathbf{r}^{\prime}\right)\right|^{3}}\label{eq:biot-savart}
\end{equation}
which can be computed for the magnetopause, Pedersen, and Hall current
systems separately \citep{rastatter14}.

\subsection{Results}

\begin{figure*}
\begin{centering}
\noindent \makebox[\textwidth]{\includegraphics{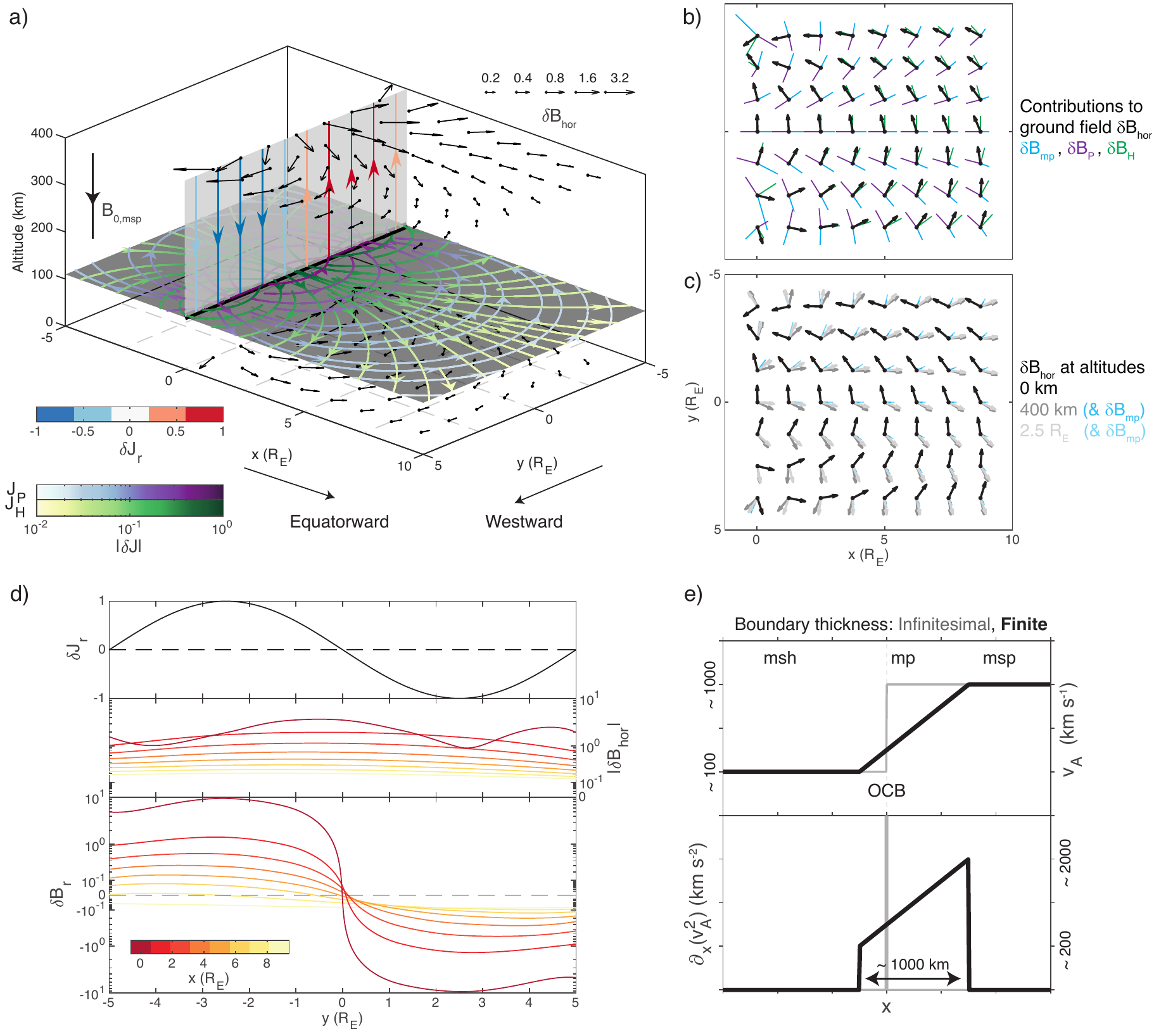}}
\par\end{centering}
\caption{Results from box model theory at the northern ionosphere. (a) Streamlines
of the perturbation magnetopause (red/blue), Pedersen (purple) and
Hall (green) currents. Horizontal magnetic perturbations on the ground
and above the ionosphere are shown as black arrows. (b) Directions
and relative magnitudes of the current system contributions to the
horizontal ground magnetic field (colours) normalised by the total
horizontal perturbation (black) at each position. (c) Horizontal directions
of the total magnetic perturbations on the ground (black) and above
the ionosphere (grey). Contribution from magnetopause currents for
the latter are shown in blue. (d) Magnetopause current variation with
azimuth (top) along with the horizontal (middle) and vertical (bottom)
ground magnetic perturbations at different equatorward distances (colours).
(e) Comparison of Alfv\'{e}n speed profiles and associated current
coupling for an infinitesimally thin (grey) and finite thickness (black)
magnetopause.\label{fig:box-model}}
\end{figure*}

The model used typical dayside field line lengths of $25\,\mathrm{R_{E}}$,
ionospheric conductances $\Sigma_{P}=\Sigma_{H}=5\,\mathrm{S}$ valid
for sunlit high-latitude regions, and zero magnetic shear across the
magnetopause. We present results for one example surface mode, a localised
perturbation with wavelength of $2\pi/k_{z}=50\,\mathrm{R_{E}}$ along
the field (fundamental mode) and $2\pi/k_{y}=10\,\mathrm{R_{E}}$
azimuthally, based on previous simulation results (A21; \citealp[hereafter A22]{archer22}).
This is applied to a single time, shown in Figure~\ref{fig:box-model}a,
since the entire pattern will propagate along $y$. The model size
is twice the dimensions of that shown, to mitigate potential edge
effects.

Figure~\ref{fig:box-model}a shows the Pedersen currents (purple)
provide current closure in the East-West direction for the FACs. This
is unlike in FLRs where closure is typically North-South \citep[e.g.][]{Greenwald1980}.
While the Pedersen currents are strongest along the OCB, due to the
finite conductivity they spread out significantly across the ionosphere
too. Their magnitudes fall off with distance from the OCB, but extend
well beyond the $1.56\,\mathrm{R_{E}}$ evanescent $e$-folding scale
of the magnetospheric signatures, given by $\left|k_{x}\right|^{-1}=\left(k_{y}^{2}+k_{z}^{2}\right)^{-\nicefrac{1}{2}}$.
Pedersen current patterns result in Hall current vortices (green)
surrounding the FAC sources/sinks at the OCB. Their sense of rotation
is clockwise for downward FACs and anticlockwise for upward FACs.
Hall current magnitudes decrease with distance identically to the
Pedersen currents, due to the conductivities used. Since ionospheric
velocities are given by the $\mathbf{E}\times\mathbf{B}$ drift, Pedersen
currents result in TCVs colocated with the Hall current vortices but
with the opposite sense of rotation. As the background magnetic field
is uniform in this model, convection speeds also fall off similarly
with distance from the OCB.

We now focus on the directions of the horizontal magnetic field perturbations,
shown in Figure~\ref{fig:box-model}a as black arrows above and below
the ionosphere. The ground magnetic signals exhibit a vortical pattern
centred at the midpoint of the FACs. Figure~\ref{fig:box-model}b
shows the relative contributions of different current systems (colours)
to the overall horizontal field (black). This reveals the field points
mainly in the direction of the Hall current contribution, with magnetopause
and Pedersen current magnetic fields largely opposing one another.
In the case of an infinite plane Alfv\'{e}n wave (for vertical background
field and uniform conductances) the ground magnetic field is entirely
dictated by Hall currents with FAC and Pedersen contributions cancelling
\citep{hughes74}. However, we notice in Figure~\ref{fig:box-model}b
that the total horizontal field is slightly misaligned from the Hall
current contribution, due to magnetopause and Pedersen contributions
no longer perfectly cancelling. This occurs because, in the case of
a surface wave, while the largest contributions to the ground field
from the Hall and Pedersen currents still arise from above the ground
station, the FACs into the ionosphere are confined to the OCB and
thus are generally not directly overhead. This results in the misalignment
growing with distance from the OCB, as well as very close to the OCB
but on the outer edges of the localised FACs, as can be seen in Figure~\ref{fig:box-model}b.

Figure~\ref{fig:box-model}a shows that the horizontal field perturbations
above and below the ionosphere are rotated from one another, seemingly
by right angles. While an exactly $90^{\circ}$ rotation would be
expected for a plane Alfv\'{e}n wave here \citep{hughes74nature,hughes74},
\citet{Kivelson1988} suggested that no rotation in the field might
occur sufficiently far from a surface wave. We find this not to be
the case for up to the $\sim6$ $e$-folding lengths shown. Panel~c
compares the directions of the ground field (black) to those at two
altitudes above the ionosphere (greys). From these it is clear this
rotation is not $90^{\circ}$ and that the rotation angle also depends
on altitude. This is because ionospheric Pedersen and Hall currents
contribute to the magnetic field above the ionosphere, as evidenced
by the blue lines which show only the field due to magnetopause currents
(less sensitive to altitude). The difference between the grey and
blue arrows grow with distance from the OCB, for similar reasons as
to which current systems are closest. Such an effect would be less
prominent with a plane Alfv\'{e}n wave since FACs permeate space
above the ionosphere. Above the ionosphere, the direction of the magnetic
field due to only magnetopause currents are much closer to $90^{\circ}$
different from the ground than the total field. The discrepancy from
a right angle is exactly that due to the non-cancellation of magnetopause
and Pedersen currents on the ground shown in panel~b. This discrepancy
will vary quantitatively with ionospheric conductivity and surface
mode wavelength.

We now consider both horizontal and vertical components of the ground
magnetic field. The horizontal field magnitudes shown in Figure~\ref{fig:box-model}d
clearly show an overall decreasing trend with distance from the OCB
(colours). The vertical component has the same sense as the FACs
into the ionosphere, i.e. reversing direction at $y=0$. The vertical
component's magnitude also decreases with distance from the OCB. Close
to the OCB the vertical component has root mean squared (RMS) values
up to $\sim3$ times larger than the horizontal component. However,
its RMS drops off much more quickly with distance, with them becoming
equal at $x\sim0.45\,\mathrm{R_{E}}$ ($k_{x}x\sim0.3$), hence is
largely negligible at further distances. The horizontal and total
magnetic field magnitudes on the ground have RMS values $1.2\text{--}2\times$
and $1\text{--}1.7\times$ those of the incident/reflected waves,
i.e. only including the magnetopause currents (when all current systems
are considered above the ionosphere the ratios become $\sim0.6\times$).
Thus despite infinitesimal latitudinal width of FACs, the ionosphere
might not screen the surface mode from the ground as suggested by
\citet{Kivelson1988}. The ionospheric screening effect goes as $\exp\left(-kh_{isp}\right)$,
where $h_{isp}$ is the ionospheric altitude, meaning that the small
latitudinal wavenumbers are not suppressed. While in the $y$-direction
there is only one wavenumber $k_{y}$ present, in the $x$-direction
the delta function in the current has equal Fourier amplitudes at
all wavenumbers, thus small total wavenumbers $k=\sqrt{k_{x}^{2}+k_{y}^{2}}$
in this superposition may be transmitted to the ground. This highlights
the need when applying the \citet{Hughes1976} formulae for non-plane
waves to use Fourier (or spherical harmonic) decomposition, rather
than simply measures of spatial amplitude extent \citep[cf.][]{ozeke09}.

Finally, we quantify the extent to which each current system controls
the total ground magnetic field through computing the Coefficient
of Determination (see \ref{sec:R2}). Since the pattern propagates,
integration in time is equivalent to spatially along $y$. This is
performed at each distance $x$ away from the OCB. On average the
Hall currents explain $92\%$ of the variance in the ground field
across all components (further statistics are given in Table~S1),
hence they still exhibit overwhelming control within the model.

We find that changing wavelengths and ionospheric conductivities in
this model lead to qualitatively similar results, but leave a full
parameterisation to future work. However, we briefly discuss the likely
effect of introducing a finite boundary thickness, which will change
the surface magnetopause currents used into volume currents. Figure~\ref{fig:box-model}e
considers a linear Alfv\'{e}n speed variation between the two half-spaces
(black) as in \citet{Chen1974}, comparing this to the infinitesimal
boundary used thus far (grey). \citet{southwood90} note that in an
azimuthally-uniform box model, FAC sources in a cold plasma are proportional
to the product of: (1) the gradient in squared Alfv\'{e}n speed;
and (2) azimuthal derivative of the compressional magnetic field signal.
The figure shows while in the infinitesimal case term~(1) is non-zero
coincident with the OCB, for a finite thickness boundary term~(1)
instead peaks at the inner edge of the magnetopause layer due to the
high magnetospheric Alfv\'{e}n speed. This is no longer coincident
with the OCB, which will be located somewhere within the boundary.
Considering term~(2), outside of the boundary layer the surface mode's
compressions must decay exponentially with distance from the magnetopause
within the two half-spaces. The sign of the compressions must also
reverse within the boundary layer itself. These two facts necessitate
that term~(2) exhibit peaks near the boundary layer edges (see also
Figure~1 of A22). Combining both terms suggests surface mode FACs
are largest near the inner edge of the magnetopause. This conclusion
holds within a box model for more complicated transitions or if thermal
effects are included \citep{itonaga00}.

\section{Global Simulation}

We now employ a global coupled MI simulation to better understand
magnetopause surface modes' potential auroral, ionospheric, and ground
magnetic signatures within a more representative environment.

\subsection{Overview}

The Space Weather Modeling Framework (SWMF; \citealp{Toth2005,Toth2012})
is used on NASA's Community Coordinated Modeling Center (CCMC). This
includes BATS-R-US (Block-Adaptive-Tree-Solarwind-Roe-Upwind-Scheme;
\citealp{Powell1999}) global MHD magnetosphere, which is run at high-resolution
($\nicefrac{1}{8}\text{--}\nicefrac{1}{16}\,\mathrm{R_{E}}$ in the
regions considered), coupled to an electrostatic thin-shell ionosphere
\citep{Ridley2004}, where the same uniform conductances as in section~\ref{sec:Box-Model-Theory}
are used. The run, which was previously presented by A21 and A22
(based on H15), simulates the magnetospheric response to an idealised
solar wind pressure pulse under northward IMF. Full setup details
are in Table~S2.

Here we summarise salient results from the BATS-R-US simulation. H15
showed, following an initial transient response, the pulse excites
damped monochromatic compressional waves near-globally with $1.8\,\mathrm{mHz}$
frequency. Amplitudes of these waves decay with distance from the
magnetopause, with a phase reversal across the boundary. The authors
concluded these oscillations could only be explained by MSE. A21 found
that across most of the dayside, magnetopause displacements showed
little azimuthal phase variation. Indeed, the surface waves are stationary
between $09\text{--}15h$ Magnetic Local Time (MLT), despite non-negligible
magnetosheath flows being present. They demonstrated this is achieved
by the time-averaged Poynting flux inside the magnetosphere surprisingly
pointing towards the subsolar point, balancing advection by the magnetosheath
flow such that the total wave energy flux is zero. Outside of this
region, waves are seeded tailward at the dayside natural frequency
and grow due to KHI. A22 reported the magnetospheric velocity polarisation's
handedness in the stationary region (Earthward of the surface mode's
turning point) was also reversed from that typically expected, which
was found only in the tailward propagating regions. While associated
magnetic field polarisations can be reversed by the field geometry
near the cusp, Earthward of each field line's apex both polarisations
are the same. A consequence of these results is MSE must have spatially
varying wavenumbers. Across the dayside the system response is large-scale
($k_{\perp}\ll k_{\parallel}$) and thus insensitive to the flow.
Since fluctuations seeded downtail from the dayside have fixed frequency
in Earth's frame, in the flanks this results in the Doppler effect
(i.e. $\omega^{\prime}=\left|\omega-\mathbf{k}\cdot\mathbf{v}_{msh}\right|\approx k_{\perp}v_{msh}$
due to the significant flow velocities) imposing shorter wavelengths
along the magnetopause of $\sim10\,\mathrm{R_{E}}$. Normal to the
magnetopause, phase fronts inside the dayside magnetosphere slowly
propagate towards the boundary ($\ll v_{A}$). The authors argued
this results from the magnetosonic dispersion relation (equation \ref{eq:magnetosonic-dispersion})
when both compressibility and damping of the surface wave are taken
into account. Finally, the surface mode is shown to couple to MHD
body waves where their eigenfrequencies match: waveguides were found
along the equatorial terminator ($x_{GSM}=0$) and outer nightside
flanks; FLRs were identified on the equatorial terminator Earthward
of the magnetopause and in the magnetotail (A21; A22).

As in those previous studies, in this paper we focus on the MSE response
(times $t>15\,\mathrm{min}$) neglecting the directly-driven transient
activity. Perturbations (represented by $\delta$'s) from equilibrium
(represented by subscript $0$'s) are extracted by subtracting $40\,\mathrm{min}$
LOESS (locally estimated scatterplot smoothing; \citealp{Cleveland1979})
filtered data, where outliers were neglected. This removed long-term
trends well during the period of interest, with spurious values occurring
only before the arrival of the pulse. For the ionosphere and ground,
coordinates employ Northward and Eastward horizontal components as
well as the vertical/radial direction. In the magnetosphere, an equivalent
field-aligned system is used. The field-aligned direction $\hat{\mathbf{B}}_{0}$
is the $90\,\mathrm{min}$ time-average of the LOESS-filtered magnetic
field. Other directions are obtained as perpendicular projections
of the local spherical polar unit vectors ($\hat{\boldsymbol{\theta}}$
for colatitude and $\hat{\boldsymbol{\phi}}$ for azimuth). Specifically,
the perpendicular Northward direction is $\hat{\boldsymbol{\mathcal{N}}}_{\perp}=-\left(\hat{\boldsymbol{\theta}}-\left[\hat{\mathbf{B}}_{0}\cdot\hat{\boldsymbol{\theta}}\right]\hat{\mathbf{B}}_{0}\right)/\left|\hat{\boldsymbol{\theta}}-\left[\hat{\mathbf{B}}_{0}\cdot\hat{\boldsymbol{\theta}}\right]\hat{\mathbf{B}}_{0}\right|$
and perpendicular Eastward is $\hat{\boldsymbol{\mathcal{E}}}_{\perp}=\left(\hat{\boldsymbol{\phi}}-\left[\hat{\mathbf{B}}_{0}\cdot\hat{\boldsymbol{\phi}}\right]\hat{\mathbf{B}}_{0}\right)/\left|\hat{\boldsymbol{\phi}}-\left[\hat{\mathbf{B}}_{0}\cdot\hat{\boldsymbol{\phi}}\right]\hat{\mathbf{B}}_{0}\right|$.
All Fourier analysis is computed between $t=15\text{--}90\,\mathrm{min}$
and limited to frequencies $1.0\text{\textendash}2.1\,\mathrm{mHz}$,
with results being integrated over this band.

\subsection{Validity}

We must assess the validity of using this simulation to qualitatively
predict auroral, ionospheric, and ground magnetic signatures of magnetopause
surface modes. We do not aim for fully quantitative predictions due
to known limitations of current global MHD codes, such as numerical
diffusion and current sheet smearing which underestimate ULF wave
effects (\citealp{Claudepierre2009}; H15). Thus only trends in wave
patterns should be interpreted, rather than absolute amplitudes of
signatures.

\begin{sidewaysfigure*}
\begin{centering}
\noindent \makebox[\textwidth]{\includegraphics[scale=0.9]{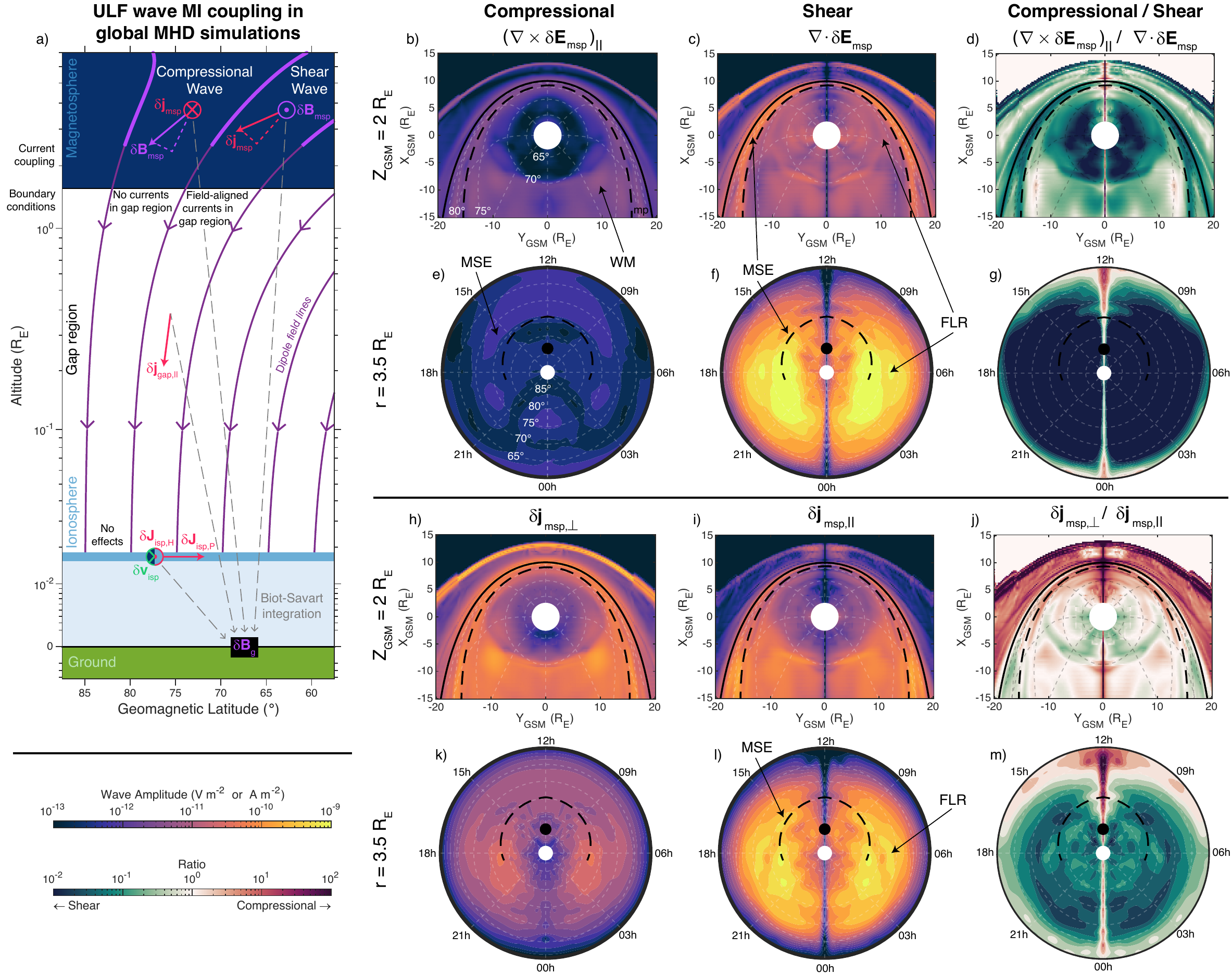}}
\par\end{centering}
\caption{(a) Illustration of magnetosphere--ionosphere coupling within global
MHD simulations applied to compressional and shear ULF waves. (b--m)
Simulation Fourier maps taken from the $z_{GSM}=2\,\mathrm{R_{E}}$
near-equatorial plane (b--d,h--j) and at $r=3.5\,\mathrm{R_{E}}$
near the simulation magnetosphere-ionosphere interface (e--g,k--m),
the latter have been projected onto the ground. These show amplitudes
of compressional (b,e) and shear (c,f) waves and associated currents
perpendicular (h,k) and parallel (i,l) to the background magnetic
field. Ratios are also shown (d,g,j,m). Solid black lines indicate
the open--closed field line boundary whereas dashed lines indicate
the magnetopause inner edge. Arrows highlight regions due to magnetopause
surface eigenmode (MSE), waveguide mode (WM), and field line resonance
(FLR).\label{fig:mi-coupling}}
\end{sidewaysfigure*}

Global MHD models are not able to simulate down to ionospheric altitudes,
e.g. since high Alfv\'{e}n speeds slow down computations, thus magnetospheric
boundary conditions are imposed on the plasma and fields further out
($r=2.5\,\mathrm{R_{E}}$ here). This leaves a gap region between
the magnetosphere inner boundary and the thin-shell ionosphere ($110\,\mathrm{km}$
altitude) which is not simulated, as depicted in Figure~\ref{fig:mi-coupling}a.
MI-coupling occurs a few grid cells radial of the magnetosphere inner
boundary ($r=3\,\mathrm{R_{E}}$), where FACs are mapped and scaled
through the gap region to the ionosphere along dipole field lines
\citep{Ridley2004}. This means equatorward of $\pm55^{\circ}$ magnetic
latitides there are no gap region FACs, hence we limit all analysis
to poleward of $60^{\circ}$. The ionosphere model solves for the
electric potential via current continuity with a given conductance
pattern, similarly to section~\ref{sec:Box-Model-Theory}, which
yields ionospheric electric fields, currents, and velocities. The
potential is mapped back to the magnetospheric inner boundary, setting
the electric field and velocity there also.

As highlighted by \citet{Kivelson1988} and indicated in Figure~\ref{fig:mi-coupling}a,
compressional and shear MHD waves will affect the ionosphere differently
due to their different currents. Shear modes exhibit FACs, hence coupling
between the magnetosphere and ionosphere will occur, which in simulations
will be performed by the current mapping. In contrast, compressional
modes have only perpendicular currents and so no current is expected
to flow between the magnetosphere and ionosphere. Purely compressional
waves result in no significant ionospheric effects, which will be
true in simulations also. Therefore, the ionospheric response to incident
ULF waves should be reliable.

Potential issues, however, arise when considering ground magnetic
field calculations. These are computed by Biot-Savart integration
of all current systems: ionospheric Pedersen and Hall currents; those
throughout the magnetosphere domain; and gap region FACs \citep{Yu2010,rastatter14}.
Because the MI-interface in simulations is much further away than
in reality, perpendicular currents at this interface will make much
smaller contributions than they would otherwise due to the increased
distance. FACs, on the other hand, are unaffected as they can traverse
the gap region. This is illustrated in Figure~\ref{fig:mi-coupling}a
and could affect both types of MHD waves, though likely more acutely
for compressional modes.

We first investigate the amplitudes of compressional and shear modes
via the curl and divergence of the electric field perturbations as
before. Due to available model outputs on the CCMC, these are calculated
via
\begin{equation}
\left(\nabla\times\delta\mathbf{E}\right)_{\parallel}=-\frac{\partial}{\partial t}\delta B_{\parallel}=i\omega\delta B_{\parallel}
\end{equation}
for the compressional mode, and 
\begin{equation}
\nabla\cdot\delta\mathbf{E}=\delta\left[\nabla\cdot\left(\mathbf{B}\times\mathbf{v}\right)\right]=\delta\left[\mu_{0}\mathbf{j}\cdot\mathbf{v}-\mathbf{B}\cdot\boldsymbol{\Omega}\right]
\end{equation}
for the shear mode, where $\boldsymbol{\Omega}=\nabla\times\mathbf{v}$
is the vorticity. Figure~\ref{fig:mi-coupling}b--d shows Fourier
wave amplitudes, along with their ratio, for a near-equatorial plane
($z_{GSM}=2\,\mathrm{R_{E}}$). At the magnetopause, the OCB is shown
as the black solid line and the magnetopause inner edge as the dashed
line, which has been manually identified based on the background current,
Alfv\'{e}n speed, and velocity polarisation (A22) and fitted to a
polynomial with local time. The large (e.g. $\sim1.5\,\mathrm{R_{E}}$
at noon) boundary width in the simulation is a consequence of MHD
being unable to resolve small gyroradius scales that dictate the $400\text{--}1000\,\mathrm{km}$
thickness of the real magnetopause \citep{Berchem1982}. Panel~b
demonstrates compressional mode amplitudes are generally largest near
the magnetopause and decay slowly across the magnetosphere with distance
from the boundary. Panel~c shows shear wave amplitudes exhibit strong
peaks inside the boundary layer, consistent with \citet{Plaschke2011},
both at the inner edge and OCB. Away from these peaks, the amplitude
falls off much more quickly than in the compressional mode. However,
shear amplitudes remain larger than compressional ones almost throughout
the equatorial magnetosphere, as indicated by the ratios in panel~d.
Since dayside FLR frequencies in the simulation are much larger than
the observed waves (A22), we conclude the large shear amplitudes are
due to non-resonant coupling between the compressional and shear modes.
This occurs due to the inhomogeneous Alfv\'{e}n speed and curvlinear
magnetic geometry present \citep[e.g.][]{Radoski1971}, resulting
in a single wave that has mixed properties of both. The same quantities
are also shown at $r=3.5\,\mathrm{R_{E}}$, near the simulation MI-interface,
in Figure~\ref{fig:mi-coupling}e--g, where these have been projected
along dipole field lines to the northern hemisphere ground. The OCB
is found to occupy a small area around the displayed black dot, indicating
a mostly closed magnetosphere as has been seen in extended northward
IMF simulations previously \citep{Song2000,Zhang2009}. The magnetopause
inner edge maps to latitudes equatorward of the OCB. Compressional
mode amplitudes (panel~e) are significantly weaker near the MI-interface,
in agreement with the expected standing structure \citep{Plaschke2011}.
They appear constrained to regions equatorward of the magnetopause
inner edge and mostly to the dayside. Shear mode amplitudes (panel~f)
exhibit strong ridges on both flanks which are well-aligned with the
inner/equatorward edge of the magnetopause, along which the amplitudes
grow with MLT away from noon. Isolated peaks also occur equatorward
of the inner edge on the terminator, corresponding to the FLRs identified
by A21. Notably no clear peak occurs at the OCB. At $r=3.5\,\mathrm{R_{E}}$
the ratio of the shear to compressional mode amplitudes are even greater
than at $z_{GSM}=2\,\mathrm{R_{E}}$ (panel~g), indicating again
the mixed properties of the wave.

The dominance of shear wave amplitudes over compressive suggests simulation
results should be reliable. However, it is currents which are more
crucial. Therefore, panels~h--m display Fourier amplitudes for the
perpendicular and parallel currents (and their ratios) at the same
locations. Reassuringly the two components have similar patterns to
the two wave modes. In particular, at $r=3.5\,\mathrm{R_{E}}$ FACs
peak along the magnetopause inner/equatorward edge and at the discrete
FLRs. No such peaks are present at the OCB, despite these occuring
out in the magnetosphere. While at $z_{GSM}=2\,\mathrm{R_{E}}$ perpendicular
and parallel currents are generally of similar magnitude, at the MI-interface
FACs dominate poleward of $\sim65\text{--}70^{\circ}$ latitudes (though
the current ratio is not as large as that for mode amplitudes). Consequently,
as perpendicular currents are small compared to FACs where MI-coupling
is performed, Biot-Savart integration will be reliable in estimating
ground magnetic field signals at high latitudes.

\subsection{Optical aurora}

FACs associated with FLR can result in periodic optical auroral forms
\citep{Greenwald1980,Samson1996,Milan2001}. Upward FACs at the ionosphere
are carried by precipitating electrons that may, if sufficiently energetic,
cause auroral emission, whereas regions of downward FACs appear relatively
darker. Given we have demonstrated surface modes also exhibit FACs,
it is worth exploring their potential auroral signatures.

As shown in panels i and l of Figure~\ref{fig:mi-coupling}, oscillatory
FACs associated with the surface mode peak at the inner edge of the
magnetopause rather than the OCB. This is difficult to intuit theoretically
within a realistic magnetosphere \citep{southwood91,itonaga00}, especially
since the simplifying assumption of wave scales being smaller than
changes in background conditions cannot be made (A22). Nonetheless,
the result is in agreement with the box model prediction of section~\ref{sec:Box-Model-Theory}.
Movie~S1 (left) shows perpendicular velocity perturbations near the
magnetospheric equator. While near the subsolar point motion is largely
normal to the boundary, away from the Sun-Earth line vortical structure
emerges near the magnetopause, particularly at the flanks. The clearest
structures have vortex cores Earthward of the OCB, corresponding to
the inner surface mode (\citealp{Lee1981}; A22). These are associated
with significant field-aligned vorticity, though this quantity is
prevalent throughout the magnetosphere (middle). In a uniform plasma
only Alfv\'{e}n waves exhibit parallel vorticity, hence this results
from non-resonant coupling between the compressional and shear modes.
On the dayside magnetosphere, the vorticity's phase structure has
shorter normal scales ($\sim6\,\mathrm{R_{E}}$) than transverse ones
(the entire morning/afternoon sector). There is also slow ($\ll v_{A}$)
phase motion towards the boundary. Since all signals' amplitudes decay
with distance from the magnetopause inner edge, the vorticity appears
to grow as its phase fronts travel towards the boundary. These features
are very similar to those reported in A21 for the compressional magnetic
field, explained as the result of surface wave damping. We also note
these boundary normal wavelengths are much larger than those expected
for field line resonance, since the large gradients in FLR frequencies
(A22) suggest scales $<1\,\mathrm{R_{E}}$ \citep{Southwood1987}.
In contrast to the dayside, tailward of approximately the terminator,
transverse wavelengths shorten to $\sim10\,\mathrm{R_{E}}$ and phase
motion appears predominantly tailward. Vorticity magnitudes in the
flanks are significantly larger than on the dayside due to KHI-amplification
and the shorter scales. We note that coupling of MSE to body modes,
such as waveguides or FLR, has been shown to occur in these regions
(A21), hence results in the nightside magnetosphere are generally
a superposition of wave modes (including surface waves). Finally,
FAC patterns (right) are very similar to the vorticity, as expected
theoretically by \citet{southwood91}. Thus, magnetopause surface
modes may exhibit FACs not just within the boundary, as in simple
box models, but throughout the magnetosphere.

Based on these results near the equatorial plane, we expect that FAC
structures at the MI-interface due to the surface mode consist of
large-scale (compared to those of Alfv\'{e}n waves) poleward moving
forms on the dayside. Indeed, this can be seen in Movie~S2 (left)
which shows the ionospheric FAC input. On the dayside, FAC latitudinal
wavelengths are $\sim10^{\circ}$ ($\sim1000\,\mathrm{km}$ in the
ionosphere), larger than expected for FLRs in this region ($<1^{\circ}$
or $100\,\mathrm{km}$). The structures propagate polewards at $1.4^{\circ}\,\mathrm{min}^{-1}$
(or equivalently $2.6\,\mathrm{km}\,\mathrm{s}^{-1}$), growing in
amplitude with their phase motion until they peak at the projection
of the magnetopause inner/equatorward edge (as demonstrated in Figure~\ref{fig:mi-coupling}i).
The azimuthal extent of these waves is likely a function of the driver,
hence solar wind excited surface waves like in the simulation should
exhibit more extended FACs than those due to most foreshock transients
\citep{Sibeck1999} or magnetosheath jets \citep{Archer2019}. In
the simulation, we find that outside of the $09\text{--}15h$ MLT
stationary region FAC structures propagate principally towards the
tail, forming periodic structure along the projection of the boundary's
inner/equatorward edge. While on the dayside structures appear azimuthally
stationary, like the surface waves in the magnetosphere, the tailward
propagating behaviour of the eigenmode outside the stationary region
causes the FACs to bifurcate at the boundary between the two regimes
during their poleward phase motion. This results in more complex structure
than simply a (spherical) harmonic wave.

We conclude that magnetopause surface modes might have optical auroral
signatures somewhat similar to FLRs. These consist of periodic brightenings
with latitudinal arc widths of $\sim5^{\circ}$ that propagate polewards
at slow speeds of $\sim1\text{--}2^{\circ}\,\mathrm{min}^{-1}$ ($\sim2\text{--}4\,\mathrm{km}\,\mathrm{s}^{-1}$).
The intensity of these periodic aurorae should amplify with their
phase motion, peaking not at the OCB as had been thought before \citep{Kozyreva2019},
but equatorward of it at the projection of the magnetopause inner
edge. In our simulation the OCB and magnetopause inner/equatorward
edge are highly separated, about $\sim7^{\circ}$ in latitude at noon.
However, this is merely due to the large magnetopause thickness. We
estimate realistic separations between the OCB and magnetopause inner/equatorward
edge to be $\sim1\text{--}2^{\circ}$ for the driving conditions considered.
This is based on the latitudinal difference in the simulation of traced
footpoint locations from field lines separated by the boundary widths
reported by \citet{Berchem1982}. Thus dayside auroral brightenings
associated with surface modes, when visible due to the time of day
/ season, might consist of poleward moving arcs that intensify towards
and peak a few degrees equatorward of an OCB proxy. Further into the
flanks, auroral brightenings may form clear periodic structure along
the projection of the boundary inner/equatorward edge, which will
principally propagate azimuthally. KHI will likely make these auroral
features generally stronger on the flanks. The auroral signatures
of surface modes might be distinguished from FLR by their higher latitude
location, lower frequency, and larger latitudinal extent. However,
we make no claims here on the character, colour, or taxonomy of such
potential auroral signatures, since these cannot be easily predicted
by MHD simulations. It also remains to be seen whether these auroral
signals can be extracted from background emissions under different
activity levels.

\subsection{Ionospheric convection}

\begin{figure*}
\begin{centering}
\noindent \makebox[\textwidth]{\includegraphics{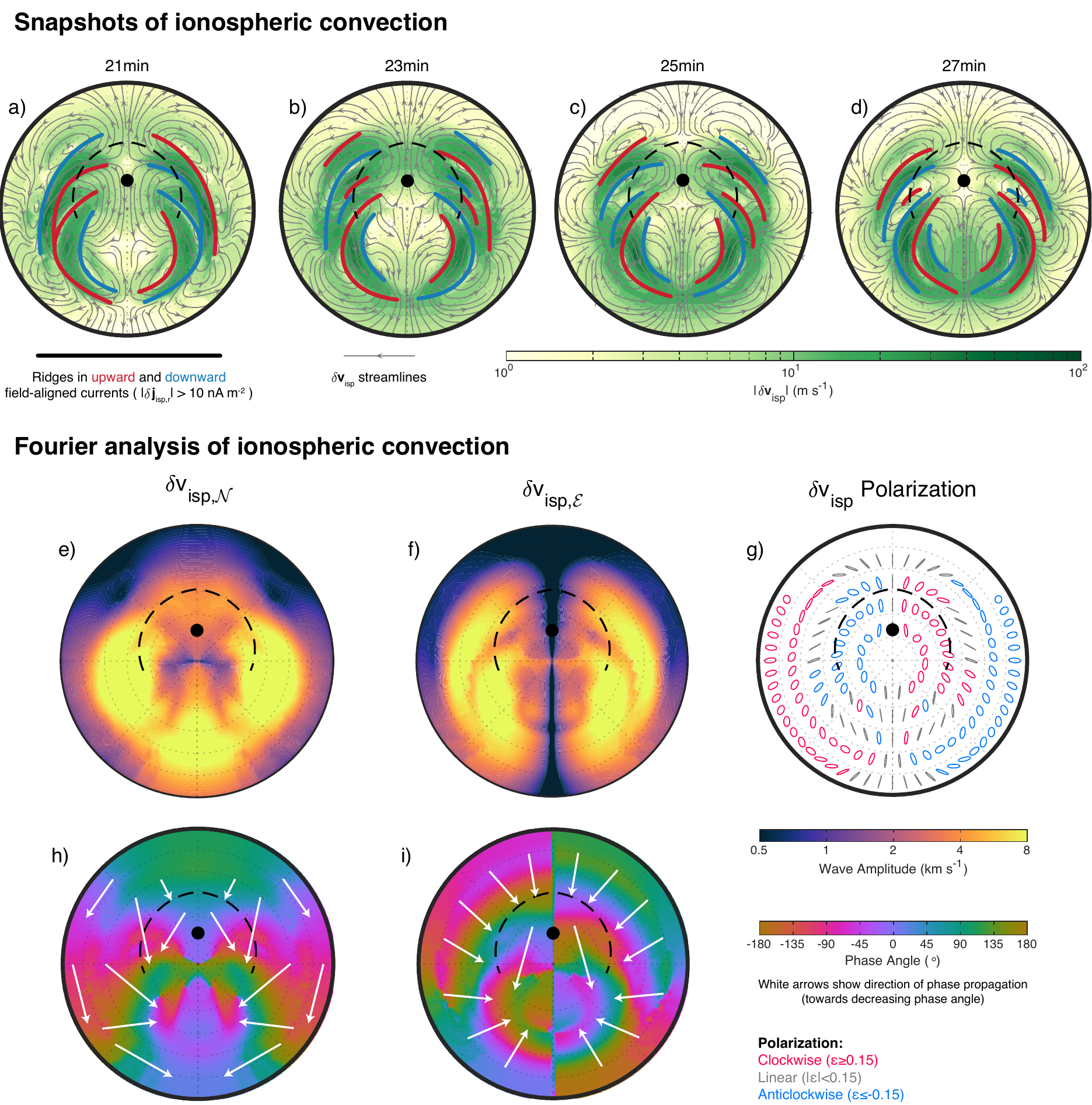}}
\par\end{centering}
\caption{(a--d) Snapshots of ionospheric convection over half an MSE cycle,
displaying the velocity magnitude (green) and streamlines (grey).
Field-aligned current ridges are also indicated. (e--i) Fourier maps,
in a similar format to Figure~\ref{fig:mi-coupling}e, of the ionospheric
velocity showing perturbation amplitudes (e,f) and phases (h,i) for
the North-South (e,h) and East-West (f,i) components. Polarization
ellipses are displayed in panel (g), coloured by the sense of rotation.\label{fig:ionosphere}}
\end{figure*}

Perturbation convection patterns are shown in Figure~\ref{fig:ionosphere}a--d
as streamline snapshots over approximately half a cycle, and in Movie~S2
as animated quivers. These reveal on the dayside large-scale convection
vortices are present, which circulate the FAC maxima (bold lines in
Figure~\ref{fig:ionosphere}a--d). Vortices are clockwise for upward
currents and anticlockwise for downward, in agreement with section~\ref{sec:Box-Model-Theory}.
Interestingly, the vortex cores are located at $\sim09h$ and $\sim15h$
MLT, i.e. the transition between stationary and propagating magnetopause
surface waves. Like the FACs, dayside vortices have shorter latitudinal
scales than longitudinal. At the lower latitudes considered though,
vortices appear more spread out in the equatorward direction. This
is likely because successive FAC structures become weaker towards
the equator, making the ionospheric response more like in the box
model. Since dayside FAC structures move polewards and grow in strength
towards the magnetopause inner/equatorward edge, the convection vortices
travel polewards and exhibit increases in speed up to this point also.
Thus poleward-moving sequences of TCVs on the dayside may be a clear
ionospheric signature of MSE. These are unlike typically reported
isolated / pairs of TCVs associated with the direct impacts of solar
wind / foreshock pressure pulses or flux transfer events, which exhibit
only tailward motion \citep{FriisChristensen1988,Sibeck1990}, suggesting
this phase motion could be a potential diagnostic for identifying
MSE in ground-based data.

Figure~\ref{fig:ionosphere}e--f shows Fourier amplitudes of the
two ionospheric velocity components. Around noon signals are mostly
North-South, like the radial motions exhibited in the magnetosphere
in this sector (A21; A22). Amplitudes peak at the magnetopause inner/equatorward
edge, just like the FACs. However, the amplitude of the East-West
component increases significantly away from noon towards the flanks.
This is most evident in panel~g, which displays polarisation ellipses
derived from the Fourier transforms (as outlined in A22). Away from
noon the ellipses' orientations rotate away from the North-South direction
and the magnitude of their ellipticity increases. Ionospheric velocity
polarisations show consistent handedness with those out in the equatorial
magnetosphere (A22), in particular a reversal is present either side
of the transition between stationary and propagating surface waves.
Panels~h--i indicate phases and propagation directions for the velocity
components, which are quite different from one another. There is little
phase variation in the North-South component on the dayside, with
only slight poleward phase motion in the stationary region. This reflects
the large-scale periodic North-South motion associated with the surface
waves that is clear from Movie~S2. In constrast, the East-West component
exhibits much larger gradients in phase latitudinally across the dayside.
These differences are due to a combination of the vortices' larger
longitudinal scales compared to latitudinal along with the poleward
motion of these vortices. Azimuthal phase variation is introduced
in the tailward propagating regime for both velocity components, though
is clearest in the North-South direction.

The finite ionospheric conductivity causes significant spreading out
of patterns caused by localised FACs. Therefore, in the above we have
focused on the dayside as we know an FLR is also present at lower
latitudes on the terminator. Figure~\ref{fig:ionosphere}e--f shows
that at the terminator two amplitude peaks on each flank are present
in both components. One of these is near the magnetopause inner boundary
corresponding to the surface mode, whereas the lower latitude peak
is due to the FLR. There is clearly significant spreading longitudinally
of the FLR-related amplitude structures by $\pm3h$, meaning that
ionospheric convection patterns in general consist of a complex superposition
of those due to the surface mode and its coupled FLR(s). Only within
$\pm3h$ of noon is the ionospheric response dominated by the surface
mode.

\begin{figure*}
\begin{centering}
\noindent \makebox[\textwidth]{\includegraphics{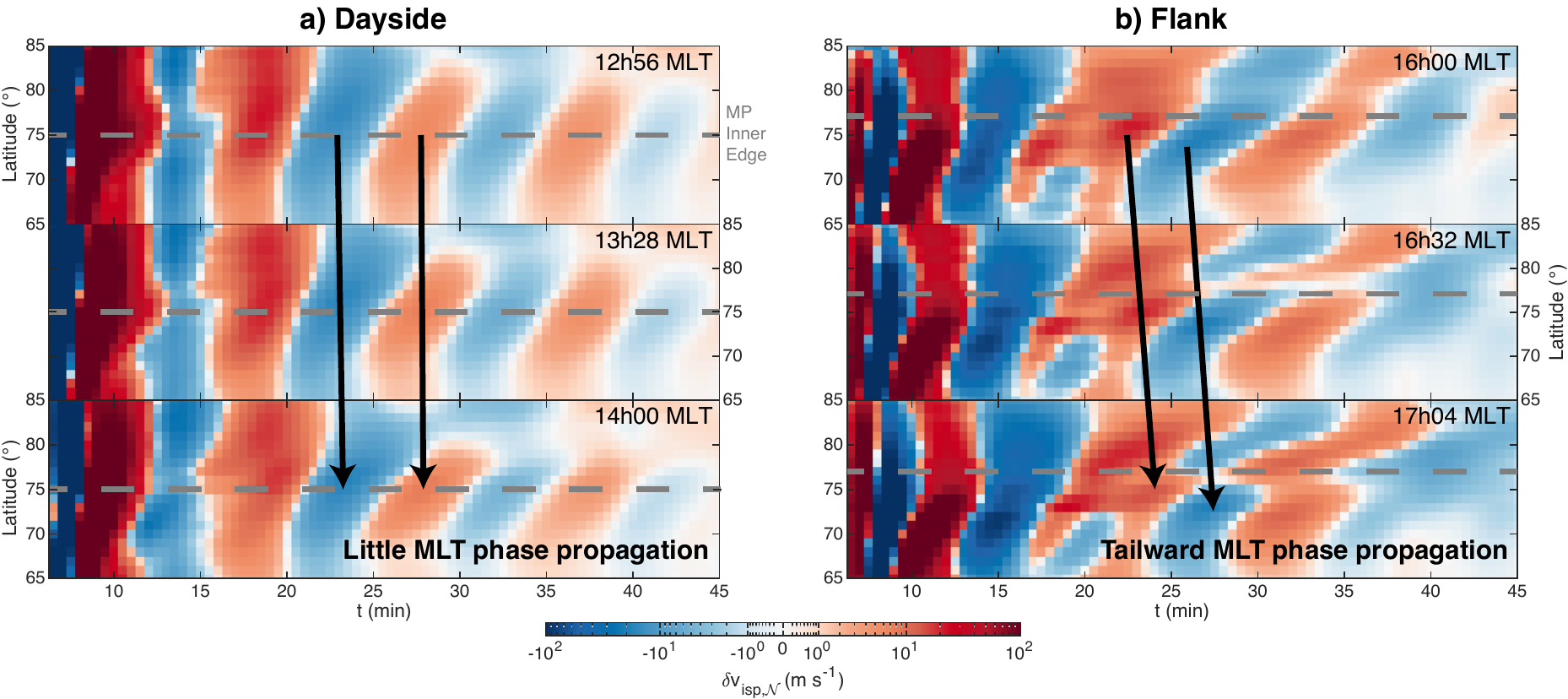}}
\par\end{centering}
\caption{Examples of potential ionospheric Doppler radar observations. Simulation
equivalents of range-time plots for adjacent beam directions (local
times) are shown for the (a) dayside and (b) flank. The projection
of the magnetopause inner edge is also indicated (grey dashed).\label{fig:superdarn}}
\end{figure*}

Figure~\ref{fig:superdarn} shows potential ionospheric Doppler radar
observations, emulating typical range-time plots. These show the time
variation of the North-South velocity perturbations with latitude
for nearby local times. Figure~\ref{fig:superdarn}a corresponds
to the dayside, which clearly shows in each panel periodic oscillations
in the ionospheric velocity that exhibit poleward phase motion and
peak in amplitude near the projection of the magnetopause inner boundary
(grey dashed line). Comparing the panels indicates there is little
phase propagation in MLT since the surface waves are stationary, again
unlike typical tailward TCVs. Figure~\ref{fig:superdarn}b corresponds
to the flanks, highlighting the increased complexity of the signal
away from noon. Nonetheless, some similar features to the dayside
are seen. While the amplitude does maximise near the magnetopause
inner boundary due to the surface mode, a secondary amplitude maximum
is present at lower latitude associated with the terminator FLR. All
of these patterns in the flank exhibit tailward phase propagation
when comparing panels.

\subsection{Ground magnetic field}

Ground magnetic field perturbations were computed using the CalcDeltaB
post-processing tool \citep{rastatter14}, performed in SM coordinates
due to the idealised model setup. We compare these to the magnetic
field signals near the MI-interface. Both are shown in Movie~S3.

\begin{figure*}
\begin{centering}
\noindent \makebox[\textwidth]{\includegraphics{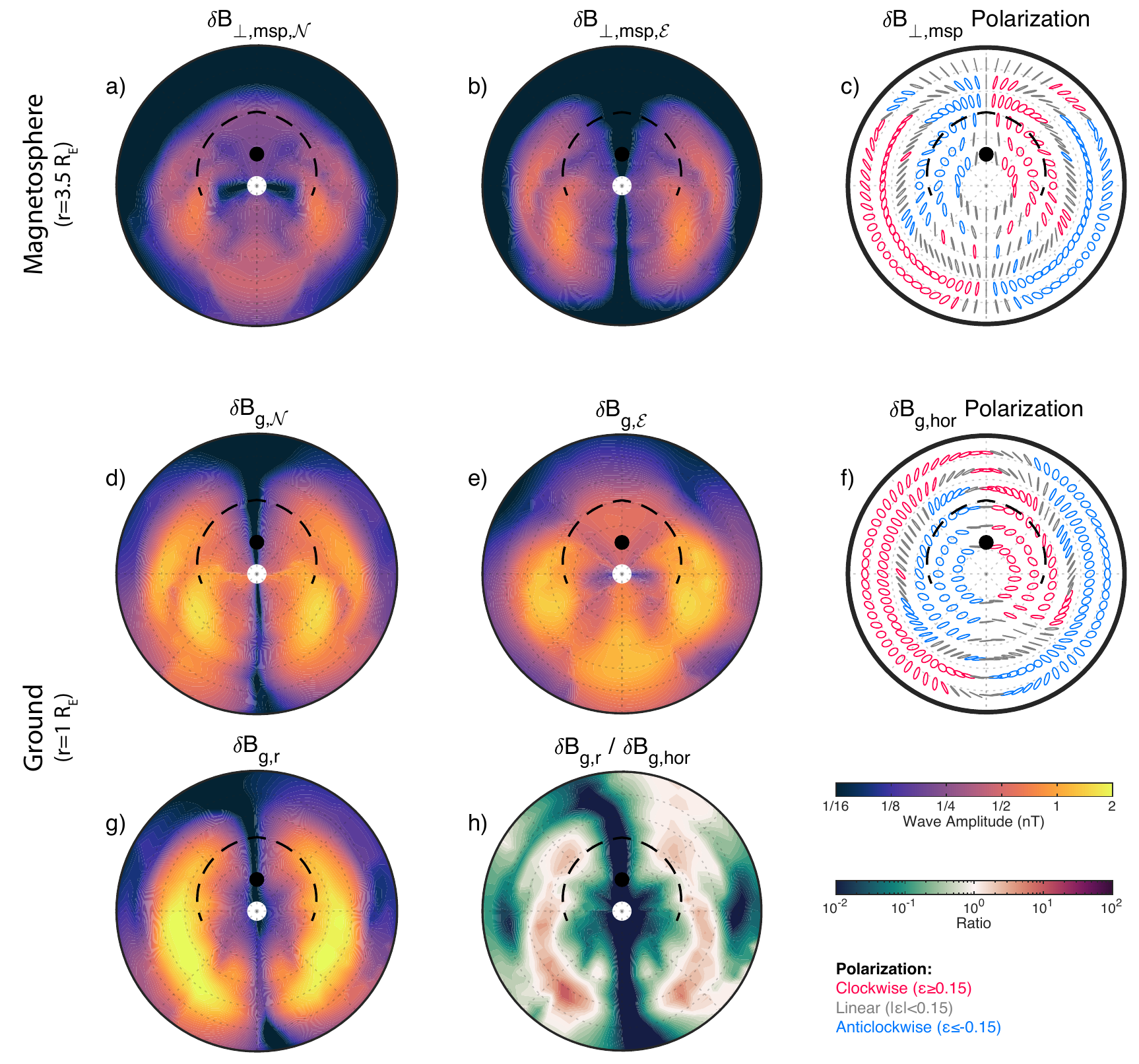}}
\par\end{centering}
\caption{Comparison of magnetic perturbations (a--c) near the magnetosphere--ionosphere
interface, and (d--h) on the ground. Fourier amplitudes (a,b,d,e,g)
and polariation ellipses (c,f) are displayed. Finally, the ratio of
vertical to horizontal ground perturbations is shown (g). Formats
are similar to Figure~\ref{fig:ionosphere}e--g.\label{fig:gmag}}
\end{figure*}

Figure~\ref{fig:gmag}a--c show the wave amplitudes and polarisations
in the magnetosphere near the MI-interface. On the dayside perturbations
are predominantly North-South oriented and maximise at the magnetopause
inner/equatorward edge. In contrast, as shown in panels~d--f, on
the ground the magnetic field is mostly in the East-West direction.
The movie shows these East-West signals are coherent across most of
the dayside, converging/diverging from the ionospheric vortex cores
at $\sim09h$ and $\sim15h$ MLT. This is unlike toroidal mode Alfv\'{e}n
waves, some of the most intensively studied ULF waves, which are aligned
mostly North-South on the ground. The Fourier amplitude maps at the
MSE frequency in Figure~\ref{fig:gmag} for the ground horizontal
components resemble those in the magnetosphere for the other component,
i.e. that at right angles. Ground signals are of significantly greater
amplitude than in the magnetosphere. While this is similar to the
box model, in the simulation this will partly be due to the scaling
of FACs across the gap region with $B_{0}$. Ground magnetic field
amplitudes also appear more extended than above the ionosphere. This
is due to the spreading of currents in the ionosphere by the finite
conductance, as discussed previously, as well as spatial integration
of these ionospheric currents \citep{Plaschke2009}. The handedness
of wave polarisations above and below the ionosphere are largely the
same. Notable differences occur at the lowest latitudes shown, where
the ground magnetic field is less reliable. Generally we see the ground
magnetic field has greater ellipticity than in the magnetosphere,
likely due to finite ionospheric conductance spreading out the currents'
vortical patterns.

The time-averaged rotation angle from the magnetosphere to the ground
was calculated for each point using the Fourier method outlined in
\ref{sec:Rotation-Angle}. Over the region depicted, this had a mean
and standard deviation of $89\pm21^{\circ}$. Note the magnetic field
near the MI-interface does not include contributions from ionospheric
Pedersen and Hall currents, hence is associated with the incident/reflected
waves only. We find in agreement with the box model that the ionosphere
rotates surface wave magnetic fields by close to $90^{\circ}$, though
significant spread in this angle occurs. Unlike in section~\ref{sec:Box-Model-Theory},
however, we found no systematic spatial ordering of the rotation angle.
This may be because in the simulation FACs are not confined to within
the boundary and move poleward. We again compute the Coefficient of
Determination at each point to quantify the contribution of different
current systems to the total ground magnetic field. Here this is done
using Fourier methods, as outlined in \ref{sec:R2}. As in the box
model, it is Hall currents which dominate the ground field, hence
why the rotation angle is close to $90^{\circ}$. However, on average
Hall currents explain only $43\%$ of the variance across all components
-- much smaller than in the simple box model. Tabel~S1 demonstrates,
however, that the other current systems (excluding Hall) and their
combinations are not significant predictors of the ground field. Therefore,
the total variance on the ground must be a complex superposition of
many current systems, including most notably Hall currents.

Movie~S3 also shows the vertical component of the ground magnetic
field. Qualitatively this somewhat resembles the FACs, in line with
predictions from the box model. However, Figure~\ref{fig:gmag}g
shows towards the flanks, unlike the FACs, the vertical field amplitudes
peak at lower latitudes than the magnetopause inner/equatorward boundary.
At the terminator the peak corresponds well with the FLRs. Therefore,
it appears that the FLR is dominating the vertical field perturbations
on the ground, relative to surface mode, across a wide local time
range. Vertical field amplitudes are generally greater than the horizontal
ones only in the vicinity of their peaks. Close to noon, however,
the vertical field is weak and only becomes significant at $\sim09h$
and $\sim15h$ MLT, the locations of dayside ionospheric vortex cores.
Thus, like with the ionosphere, only in the vicinity of noon are the
ground magnetic perturbations solely due to the surface mode.

\begin{figure*}
\begin{centering}
\noindent \makebox[\textwidth]{\includegraphics{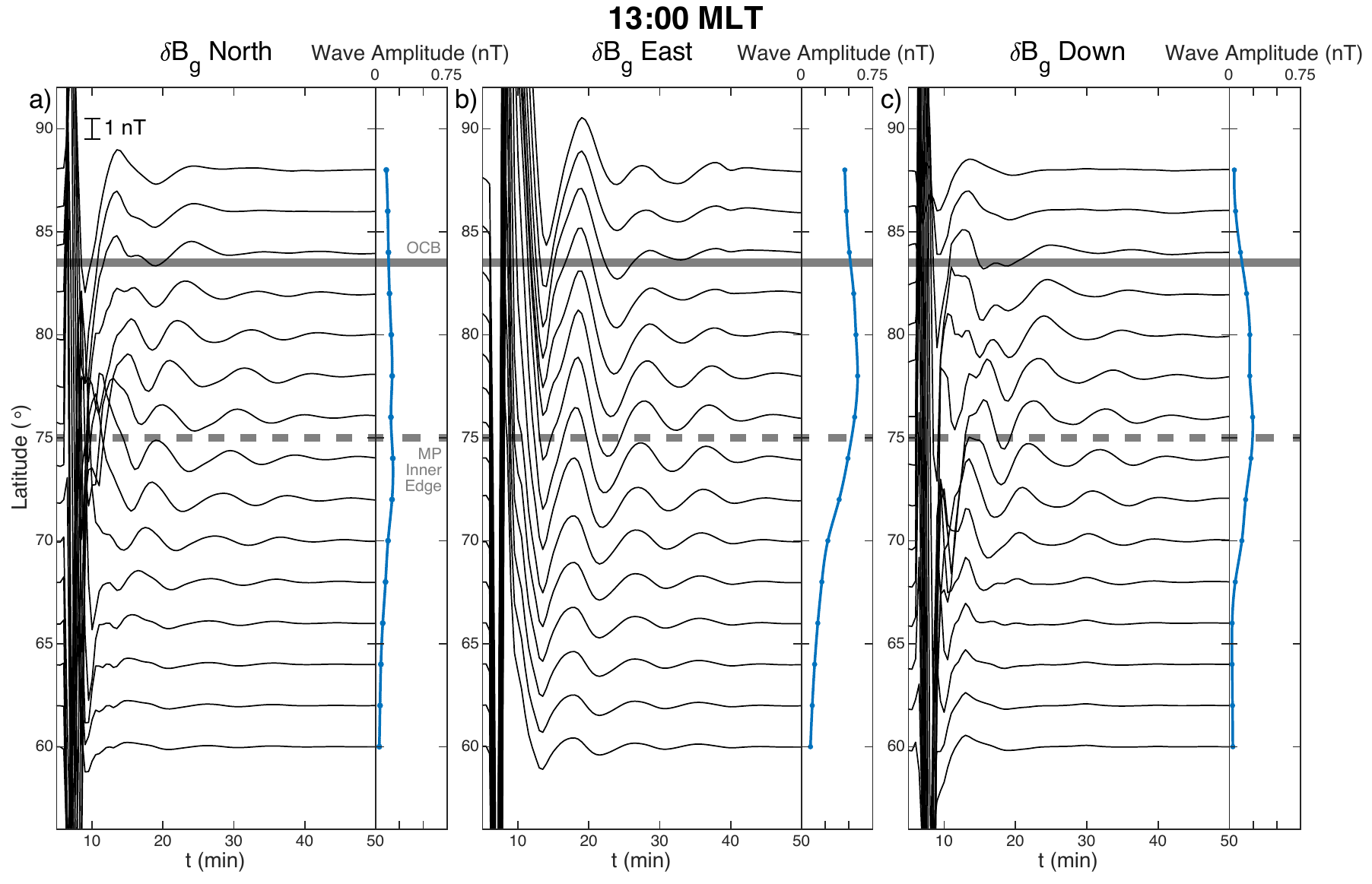}}
\par\end{centering}
\caption{Examples of potential latitudinal ground magnetometer chain observations.
Magnetic field perturbations in the (a) North-South, (b) East-West,
and (c) vertical components are shown as stacked time-series (black)
as well as Fourier amplitudes (blue). Grey lines depict projections
of the open-closed boundary (solid) and magnetopause inner edge (dashed).\label{fig:mag-chain}}
\end{figure*}

Figure~\ref{fig:mag-chain} shows stacked time-series of a latitudinal
chain of virtual ground magnetometers located close to noon, where
effects of the FLR are small. These demonstrate poleward phase motion
of the ground magnetic field in all three components, unlike ground
magnetometer observations of typical isolated / pairs of TCVs which
show predominantly tailward motion \citep[e.g.][]{FriisChristensen1988}.
Like with the ionospheric velocity though, this phase motion is slightly
different for all three components. The amplitude variation (blue
lines) is quite broad for all three components. While the vertical
component appears to peak at the magnetopause inner/equatorward edge,
the East-West component's maximum appears shifted slightly poleward
of this location and the North-South component has a rather flat peak.
Nonetheless, all three maxima are clearly closer to the magnetopause
inner/equatorward edge than the OCB within the simulation. We suggest
that all these features could be used as diagnostics for identifying
surface modes in ground magnetometer networks.

\section{Discussion}

\subsection{Limitations}

In both aspects of this study we have employed uniform ionospheric
conductances. This was to understand the surface mode's ground-based
signatures in the simplest case. Improved empirical conductance maps
typically include effects of solar illumination and/or auroral oval
conductance contributions \citep{Ridley2004}. The former exhibit
relatively small variations over scales much larger than surface mode
wavelengths, hence likely have little effect on the predictions. In
contrast, auroral oval Hall conductances can be significantly larger
than those outside this region. While these could result in stronger
currents within the auroral oval, and thus stronger ground magnetic
signals, the sense of FAC closure would likely remain. \citet{Hartinger2017}
performed simulations comparing ground field perturbations from a
solar wind pressure increase under different conductance models. They
found qualitatively similar results for the uniform and solar conductance
patterns, but somewhat different amplitude profiles with the auroral
pattern.

So far we have treated the ground as a perfect insulator, in line
with most past global modelling and observational work \citep[e.g.][]{Samsanov2015,Tanaka2020}.
While some other ULF wave studies have considered the ground to be
a perfect conductor \citep[e.g.][]{hughes74,hughes74nature,Waters2008},
neither regime realistically includes contributions from induced telluric
currents in the ground. To estimate their likely effect we apply the
Complex Image Method \citep{Boteler1998,Pirjola1998}. This places
an image current, with the same strength as that in the overhead ionosphere,
at a depth of $h_{isp}+2p$ for complex skin depth $p$. Here we take
this skin depth to be
\begin{equation}
p=\frac{1}{\sqrt{i\omega\mu_{0}\sigma}}
\end{equation}
where $\sigma$ is the ground conductivity. One can assume a uniform
conductivity that is able to capture the spatial variations in the
field/currents through a plane wave \citep{Pirjola2009}. For wavelengths
$>200\,\mathrm{km}$ and periods $<1\,\mathrm{min}$, valid for MSE,
this yields values of $\sim1\text{--}2\,\mathrm{mS}\,\mathrm{m}^{-1}$,
in line with ground conductivities for rocky or city areas \citep{cebik2001}.
These values along with the simulation MSE frequency result in skin
depths of magnitude $190\text{--}270\,\mathrm{km}$, much greater
than the ionospheric altitude. Given that in both the box model and
simulation the ground magnetic field was mostly dictated by Hall currents
with large scale sizes, we estimate the ground magnetic fields from
telluric currents by assuming an infinite line current in the ionosphere
\citep{Boteler1998}. This predicts horizontal ground magnetic field
perturbations are amplified by $18\text{--}24\%$ and vertical fields
reduced by $3\text{--}6\%$ due to the induced ground currents. Phase
changes are negligible ($<7^{\circ}$). These are relatively small
contributions due to the low $1.8\,\mathrm{mHz}$ frequency of the
surface mode, since lower frequencies are less effective at inducing
telluric currents for a given amplitude. In contrast, higher frequency
ULF waves of $10\text{--}100\,\mathrm{mHz}$ are predicted to change
the horizontal field by $40\text{--}70\%$. We also note that near
oceans, the high $5\,\mathrm{S}\,\mathrm{m}^{-1}$ conductivity of
salt water likely renders all ULF waves' ground signatures greatly
affected ($>90\%$). Future studies could more comprehensively investigate
the importance of telluric currents to ground magnetometer signals
of surface and other ULF waves.

Our brief evaluation of these limitations suggests the use of a wide
range of latitudes when examining potential observations to these
predictions, where overall trends likely persist. On smaller scales,
local effects due to varying conditions in the ionosphere and the
ground may be more important, which could form the basis of future
work.

\subsection{Comparison to previous observations}

Auroral brightenings, ionospheric convection vortices, and ground
pulsations are expected from FACs in general. However, mapping observations
from the ground out to space is difficult when trying to distinguish
surface waves from nearby body waves. We thus limit comparitive discussion
to studies that could better constrain ground-based observations.

Previous conjugate observations have linked aurorae to surface waves.
\citet{He2020} suggested sawtooth aurorae, large-scale undulations
along the equatorward edge of the diffuse aurora, may be the optical
atmospheric manifestation of plasmapause surface waves. During a geomagnetic
storm, they showed $1.4\,\mathrm{mHz}$ plasmapause surface waves
(i.e. with similar frequencies to MSE) in the afternoon-dusk correlated
with sawtooth auroral patterns near the footpoints of the plasmapause
field lines, with wavelengths and propagation speeds of both being
in agreement. The authors suggested the modulation of hot plasma by
the plasmapause surface mode may have led to particle precipitation,
and thus diffuse aurora. Similarly, \citet{Horvath2021} presented
two case studies of KHI-waves on the flank magnetopause during geomagnetic
storms, which appeared to excite surface waves on / near the plasmapause
in the hot zone of the outer plasmasphere. During these events, correlated
complicated sub-auroral plasma flows and large auroral undulations
were observed at low Earth orbit. The authors conclude magnetopause
surface modes couple, in complex ways, to the inner magnetosphere
and auroral zones. Finally, the ground-based study of \citet{Kozyreva2019}
used observations of the equatorward edge of the red cusp aurora as
an optical proxy for the OCB following southward IMF turnings. They
noted quasi-periodic motion of this boundary in 3~events, which they
interpreted as evidence for MSE. While these studies demonstrate a
connection between magnetopause/plasmapause surface waves and aurorae,
the latter's generation mechanism remains unclear. Further work should
determine whether surface waves' FACs directly result in periodic
aurorae or merely perturb existing auroral current systems.

\citet{Archer2019} first showed MSE signatures may be present in
dayside ground magnetometer data, though unfortunately data was low
resolution and had poor spatial coverage which limited conclusions.
\citet{Kozyreva2019} presented data from a near-noon latitudinal
chain following impulsive external driving. Short-lived oscillations
in the North-South component were found to peak $\sim1\text{--}2^{\circ}$
equatorward of the optical OCB proxy. While the authors attributed
this to experimental uncertainty, it is unclear why this would result
in a systematic effect. The offset agrees with our estimates for the
inner/equatorward edge of the magnetopause boundary layer, thus could
instead be consistent with our results. They also noted poleward phase
motion and large $\sim8\text{--}10^{\circ}$ latitudinal scales, both
like in our simulation. \citet{He2020} demonstrated ground magnetic
pulsations associated with plasmapause surface waves in the afternoon-dusk
sector. A latitudinal magnetometer chain showed clear poleward phase
motion, as in the simulation. However, the authors suggest the plasmapause
surface wave may have coupled to an FLR outside the plasmasphere due
to the amplitude and phase structure observed, potentially complicating
these observations. Finally, \citet{Hwang2022} presented two case
studies of KHI-waves and their ground effects. Vortical horizontal
field perturbations were observed at high latitudes, corresponding
to bead-like FACs elongated in the east-west direction like those
seen in the simulation flanks.

There are some interesting similarities and differences between these
magnetopause surface mode results and the classic Sudden Commencement
(SC) or, more generally, TCV response of the magnetosphere. These
models are generally used to interpret ground-based observations following
impulsive solar wind driving, e.g. interplanetary shocks or solar
wind pressure pulses, hence warrant further discussion. The \citet{Araki1994}
model of SC predicts bipolar variations of the geomagnetic field at
polar latitudes due to a global ionospheric twin vortex resulting
from pairs of FACs on each of the morning and afternoon sectors. These
are similar but larger in scale than typical TCVs. Our simulation
results are broadly consistent with this model during the transient
period (see Movie~S2 and Figure~6). The \citet{Araki1994} model,
however, does not predict the subsequent periodic oscillations following
this transient, which are associated with MSE. This is because it
only considers the intensification of magnetopause currents and FACs
related to a single ripple on the magnetopause, hence not a surface
wave or eigenmode. This is also true of similar TCVs models \citep[e.g.][]{Sibeck1990}.
MHD wave propagation during SC is typically linked to the \citet{Tamao1964a,Tamao1964b}
path or cavity/waveguide theory \citep{Kivelson.1984,Kivelson1985},
where compressional waves couple to FLRs at the location their (eigen)frequencies
match. Thus the possible contributions of magnetopause surface waves
has not been considered in past SC observations or modelling \citep[see the review of][]{Fujita2019}.
SC can often be followed by long $\geq10\,\mathrm{min}$ period waves
\citep[e.g.][]{Matsushita1962}, however, these are rarely discussed.
It is therefore possible that ground-based evidence of MSE could be
prevalent in past SC observations, with the subsequent pulsations
either being neglected or misidentified as cavity/waveguide modes
and FLR, which are generally not expected to occupy such low frequencies
on the dayside \citep{Archer2015a}.

Overall, our simulation results appear consistent with the few previous
reported observational signatures of magnetopause surface waves specifically.
However, some aspects could not be tested with the data presented,
motivating the need for both dedicated future observational studies
and reanalysis of previously examined events in light of this work.

\subsection{Implications}

The results presented in this paper have potential consequences within
the context of space weather, which we briefly comment on.

We have demonstrated magnetopause surface modes are predicted to result
in ionospheric currents and electric fields. This offers the possibility
that surface wave energy may be dissipated in the ionosphere, like
it is for other ULF wave modes \citep[e.g.][]{Glassmeier1984}. Joule
heating rates are given by

\begin{equation}
\begin{array}{rl}
\mathbf{j}\cdot\mathbf{E} & =\frac{1}{\Sigma_{P}}E^{2}\\
 & =\frac{1}{\Sigma_{P}}\left(E_{0}^{2}+2\mathbf{E}_{0}\cdot\delta\mathbf{E}+\delta E^{2}\right)
\end{array}
\end{equation}
where the first term corresponds to the DC equilibrium heating rate
and the subsequent terms are associated with pulsations. Recall a
uniform Pedersen conductance of $5\,\mathrm{S}$ was used, which is
reasonable for sunlit high-latitude regions \citep[cf.][]{Ridley2004}.
In the simulation, we integrate these over the entire dayside ionosphere.
While the DC rate is $3\,\mathrm{GW}$, we find the maximum pulsation-related
rate is $40\,\mathrm{GW}$, occurring during the transient response.
The peak dissipation rate during times of confirmed MSE ($t>15\,\mathrm{min}$)
is also significant compared to the background at $0.4\,\mathrm{GW}$
(i.e. 13\%). While the inclusion of the KHI-amplification of the surface
waves and their coupling to FLRs on the nightside result in the global
ionospheric dissipation rates being even greater during MSE times
at 25\% of the global DC rate, the ionospheric conductances in the
nightside hemisphere are less realistic.  Overall, these simulation
results qualitatively suggest magnetopause surface modes may provide
important contributions to ionospheric heating. Further work could
quantitatively predict heating rates due to surface waves using a
range of more representative wave amplitudes and ionospheric conditions,
improving our understanding of their global significance under different
driving regimes.

We also predict that magnetopause surface modes result in oscillatory
magnetic fields at Earth's surface. This suggests they could be a
source of geomagnetically induced currents driven by geoelectric fields
\citep[e.g.][]{Belakhovsky2019,Heyns2021}. While geoelectric fields
are frequency-dependent with a higher frequency bias relative to the
underlying disturbance geomagnetic field \citep[e.g.][]{Boteler1998,Pirjola1998},
distinct Pc5-6 frequency ULF waves ($1\text{--}7\,\mathrm{mHz}$)
can result in significant measured geoelectric fields \citep{Hartinger2020,Shi2022}.
Therefore, it may be possible that magnetopause surface waves, at
either MSE or higher (e.g. intrinsic KHI) frequencies, could similarly
result in strong geoelectric fields. Modelling this is beyond the
scope of this study, as it is known the three-dimensional conductivity
structure of Earth is important in accurate characterisation of geoelectric
fields \citep{Bedrosian2015}. Therefore, further study is warranted
in assessing whether magnetopause surface modes of more realistic
amplitudes may be a source of geoelectric fields and thus geomagnetically
induced currents \citep[cf.][]{Belakhovsky2019,Heyns2021,Yagova2021}.

\section{Conclusions}

We have investigated magnetopause surface waves' direct effects on
the aurorae, ionosphere, and ground magnetic field through both MHD
theory and a global MI-coupling simulation. Our main conclusions are
as follows:
\begin{enumerate}
\item Magnetopause surface modes have associated FACs into / out of the
ionosphere, which for a finite thickness boundary maximise at the
inner/equatorward edge of the magnetopause rather than the OCB. Non-resonant
coupling between the compressional and Alfv\'{e}n modes results in
further monochromatic FACs throughout the magnetosphere, hence are
unrelated to the Alfv\'{e}n continuum. The amplitudes of these currents
fall off with distance from the magnetopause.
\item The normal phase structure reported by A21 in the equatorial magnetosphere
due to damping also manifests at the MI-interface as slow $\sim1\text{--}2^{\circ}\,\mathrm{min}^{-1}$
poleward moving FAC structures. With latitudinal wavelengths of $\sim10^{\circ}$
on the dayside, these are large-scale compared to expectations for
FLR.
\item FACs associated with global MSE are weakest on the dayside, due to
smaller boundary displacements and azimuthal scales that span the
morning/afternoon sector. In the flanks, where the Doppler effect
imposes shorter scales and wave perturbations are amplified through
KHI, strong periodic FACs may be present along the magnetopause inner/equatorward
edge.
\item We suggest upward FACs associated with surface modes may lead to periodic
auroral brightenings that peak in intensity at the magnetopause inner/equatorward
edge. On the dayside, these auroral forms move slowly poleward at
$\sim1\text{--}2^{\circ}\,\mathrm{min}^{-1}$ ($\sim2\text{--}4\,\mathrm{km}\,\mathrm{s}^{-1}$)
and occupy large $\sim5^{\circ}$ latitudinal bands compared to narrower
FLR-related aurorae. Periodic longitudinal structure will reflect
that out at the magnetopause..
\item Ionospheric convection vortices circulate the surface mode's FAC structures.
Like the FACs, they are large-scale, poleward moving, and strongest
at the magnetopause inner/equatorward edge. The finite conductivity
causes ionospheric signals to be more spread out than the FACs.
\item Magnetopause surface modes can also cause ground magnetic field signals.
These are largely caused by Hall current vortices, which rotate the
magnetic field perturbations from above the ionosphere to the ground
by almost $90^{\circ}$ (though significant non-systematic spread
in this rotation angle occurs). Therefore, ground signatures of MSE
near noon are strongest in the East-West direction. Oscillations have
the same frequency across $L$-shells, amplitudes peak near the magnetopause
inner/equatorward edge, and latitudinal variations are large-scale.
\end{enumerate}
These conclusions provide qualitative predictions for magnetopause
surface modes which might be applied to interpreting high-latitude
ground-based data

Different plasma conditions throughout geospace are expected to slightly
modify some of the features observed. Penetration of surface modes
into the magnetosphere is governed by local normal wavenumbers (equation~\ref{eq:magnetosonic-dispersion}).
This is very weakly dependent on the Alfv\'{e}n speed distribution
for MSE, due to its such low frequencies compared to FLR, thus varying
the simulation inner boundary density or including a ring current
would have little effect. The inclusion of a plasmasphere \citep{Claudepierre2016},
however, might have a greater influence since it could lower the Alfv\'{e}n
continuum at the plasmapause to MSE-range frequencies. This would
limit the surface mode's access to the inner magnetosphere and lead
to coupling to plasmapause surface waves and/or FLR on the dayside,
introducing an additional superposition of waves. Nonetheless, surface
mode signatures in the outer magnetosphere would be little affected.
Different Alfv\'{e}n speed distributions would, however, change the
eigenfrequencies of WMs and FLRs, hence where MSE may excite these
secondary waves. The polar cap in the simulation presented is very
small. While our results (both theoretically and in the simulation)
suggest surface mode currents at the MI-interface maximise at the
inner/equatorward edge of the magnetopause boundary rather than the
OCB, whether the amount of open flux affects possible currents near
the OCB should be explored in future work. We also note this simulation
is North-South and dawn-dusk symmetric due to zero dipole tilt, whereas
interhemispheric \citep[e.g.][]{Engebretson2020,Xu2020} and dawn-dusk
\citep[e.g.][]{Henry2017} asymmetries in geospace are of great interest
in understanding solar wind -- magnetosphere -- ionosphere coupling.
Overall, how magnetopause surface modes manifest in the magnetosphere,
ionosphere, and on the ground under varying magnetospheric conditions
should be ascertained through series of further simulations.

Finally, a further potential avenue for remote sensing magnetopause
(or even plasmapause) surface modes could be ionospheric Total Electron
Content (TEC) data derived from Global Navigation Satellite Systems.
Recent work has shown that ULF waves may modulate TEC, yielding periodic
oscillations of similar frequency \citep{Pilipenko2014,Watson2015,Balakhovsky2016,Kozyreva2020,Zhai2021}.
While several mechanisms for this have been proposed, overall they
remain poorly understood. Future work could use the predicted ionospheric
electric fields associated with surface modes to drive height-resolved
ionosphere and neutral atmosphere models \citep{Ozturk2020}. This
would crucially unveil how the coupled ionosphere -- thermosphere
-- mesosphere system reacts to boundary waves in our magnetosphere.

\clearpage{}


\clearpage{}

\appendix

\section{Coefficient of Determination\label{sec:R2}}

The Coefficient of Determination, $R^{2}$, is the proportion of the
variance in a dependent variable that is predictable from an independent
variable, often used in statistical models to quantify how well outcomes
are replicated. In the case that the variables are complex vector
time-series, this is given by
\begin{equation}
R^{2}=1-\frac{\int dt\left(\mathbf{y}-\mathbf{f}\right)\cdot\left(\mathbf{y}-\mathbf{f}\right)^{*}}{\int dt\mathbf{y}\cdot\mathbf{y}^{*}}
\end{equation}
where $\mathbf{y}$ is the dependent variable and $\mathbf{f}$ is
its predicted/modelled value. It follows from Parseval's theorem that
for oscillatory signals $R^{2}$ can also be determined from the complex
Fourier amplitudes (denoted by tildes here)
\begin{equation}
R^{2}=1-\frac{\int df\left(\tilde{\mathbf{y}}\cdot\tilde{\mathbf{y}}^{*}-\tilde{\mathbf{y}}\cdot\tilde{\mathbf{f}}^{*}-\tilde{\mathbf{f}}\cdot\tilde{\mathbf{y}}^{*}+\tilde{\mathbf{f}}\cdot\tilde{\mathbf{f}}^{*}\right)}{\int df\tilde{\mathbf{y}}\cdot\tilde{\mathbf{y}}^{*}}
\end{equation}
$R^{2}$ can be interpreted as the fraction of explained variance,
with $R^{2}=1$ corresponding to perfect prediction. Negative values
are possible when the predictor performs worse than one which always
predicts the mean value. In this paper we use $R^{2}$ to indicate
what proportion of the total ground magnetic field perturbations are
determined by contributions from different current systems. This has
the benefits over simply comparing the root-mean-squared magnitudes
of contributions \citep[e.g.][]{rastatter14} since the vector directions
are also included.

\section{Rotation Angle\label{sec:Rotation-Angle}}

The signed right-handed rotation angle from a vector $\mathbf{b}$
to $\mathbf{a}$ about some direction $\hat{\mathbf{z}}$ may be expressed
as
\begin{equation}
\Delta\theta=\mathrm{atan2}\left(\left[\mathbf{b}\times\mathbf{a}\right]\cdot\hat{\mathbf{z}}\,,\,\mathbf{b}\cdot\mathbf{a}\right)
\end{equation}
In the case of oscillatory vectors, the time-average (angular brackets)
of $\Delta\theta$ can be arrived at by using the following properties
of complex Fourier amplitudes

\begin{equation}
\begin{array}{ccc}
\left\langle \mathbf{b}\times\mathbf{a}\right\rangle  & = & \frac{1}{2}\mathrm{Re\left\{ \tilde{\mathbf{b}}\times\tilde{\mathbf{a}}^{*}\right\} }\\
\left\langle \mathbf{b}\cdot\mathbf{a}\right\rangle  & = & \frac{1}{2}\mathrm{Re\left\{ \tilde{\mathbf{b}}\cdot\tilde{\mathbf{a}}^{*}\right\} }
\end{array}
\end{equation}
and hence
\begin{equation}
\left\langle \Delta\theta\right\rangle =\mathrm{atan2}\left(\mathrm{Re}\left\{ \tilde{\mathbf{b}}\times\tilde{\mathbf{a}}^{*}\right\} \cdot\hat{\mathbf{z}}\,,\,\mathrm{Re}\left\{ \tilde{\mathbf{b}}\cdot\tilde{\mathbf{a}}^{*}\right\} \right)
\end{equation}
In this paper, we use this to calculate the rotation of the horizontal
magnetic field components from above the ionosphere to below it. Thus
the direction $\hat{\mathbf{z}}$ is taken as the vertical/radial.

\begin{notation}

\notation{$msh$}Magnetosheath

\notation{$mp$}Magnetopause

\notation{$msp$}Magnetosphere

\notation{$isp$}Ionosphere

\notation{$g$}Ground

\notation{$GSM$}Geocentric Solar Magnetospheric coordinates

\notation{$SM$}Solar Magnetic coordinates

\notation{$\mathcal{N}$}Northward component

\notation{$\mathcal{E}$}Eastward component

\notation{$hor$}Horizontal

\notation{$\parallel$}Parallel to the magnetic field

\notation{$\perp$}Perpendicular to the magnetic field

\notation{$n$}Normal to boundary

\notation{$P$}Pedersen

\notation{$H$}Hall

\notation{$\epsilon$}Ellipticity

\notation{$\mu_{0}$}Vacuum permeability

\notation{$\sigma$}Conductivity

\notation{$\Sigma$}Conductance

\notation{$\psi$}Electrostatic potential

\notation{$\omega$}Angular frequency

\notation{$\Omega$}Vorticity

\notation{$\mathbf{B}$}Magnetic field

\notation{$c_{s}$}Speed of sound

\notation{$\mathbf{E}$}Electric field

\notation{$h$}Altitude

\notation{$\mathbf{j}$}Current density

\notation{$\mathbf{k}$}Wave vector

\notation{$L$}McIlwain field-line equatorial distance parameter

\notation{$\mathbf{r}$}Geocentric Position

\notation{$s$}Field line length

\notation{$t$}Time

\notation{$\mathbf{v}$}Plasma velocity

\notation{$v_{A}$}Alfv\'{e}n speed

\end{notation}

\section*{Open Research}

Simulation results have been provided by the Community Coordinated
Modeling Center (CCMC) at Goddard Space Flight Center using the SWMF
and BATS-R-US tools developed at the University of Michigan's Center
for Space Environment Modeling (CSEM). This data is available at \url{https://ccmc.gsfc.nasa.gov/results/viewrun.php?domain=GM&runnumber=Michael_Hartinger_061418_1}
and \url{https://ccmc.gsfc.nasa.gov/results/viewrun.php?domain=PP&runnumber=Martin_Archer_20211219_PP_1}.
Box model results generated are available in \citet{archer_data}.
\begin{acknowledgments}
We acknowledge valuable discussions within the ISSI International
Team project \#546 ``Magnetohydrodynamic Surface Waves at Earth's
Magnetosphere (and Beyond)'', led by MOA and Katariina Nykyri, thanks
to support by the International Space Science Institute (ISSI) in
Bern. MOA holds a UKRI (STFC / EPSRC) Stephen Hawking Fellowship EP/T01735X/1.
MDH was supported by NASA grant 80NSSC19K0907 and NSF grant AGS 2027210.
DJS was supported by STFC grant ST/S000364/1. MH was supported by
NERC grant NE/V003070/1 and Schmidt Science Fellows, in partnership
with the Rhodes Trust. JWBE was supported by NERC grants NE/P017142/1
and NE/V003070/1. A.N.W. was partially funded by STFC grant ST/N000609/1.
XS is supported by NASA award 80NSSC21K1677 and NSF award AGS-1935110.
For the purpose of open access, the author has applied a Creative
Commons Attribution (CC BY) licence to any Author Accepted Manuscript
version arising.
\end{acknowledgments}

\bibliography{mse_ground}

\end{document}